\documentclass[sn-mathphys,Numbered]{sn-jnl}
\pdfoutput=1

\usepackage{tabularx}%
\usepackage{chngcntr}     

\usepackage{appendix}
\usepackage{float}
\usepackage{newfloat}%
\DeclareFloatingEnvironment[fileext=pdf,name=Extended Data Figure]{extfigure}%

\usepackage{graphicx}%
\usepackage{multirow}%
\usepackage{amsmath,amssymb,amsfonts}%
\usepackage{amsthm}%
\usepackage{mathrsfs}%
\usepackage[title]{appendix}%
\usepackage{xcolor}%
\usepackage{textcomp}%
\usepackage{manyfoot}%
\usepackage{booktabs}%
\usepackage{algorithm}%
\usepackage{algorithmicx}%
\usepackage{algpseudocode}%
\usepackage{listings}%
\usepackage{upgreek}%
\usepackage{comment}%
\usepackage{ulem} 

\newcommand{\delete}[1]{\sout{#1}}
\newcommand{\replace}[2]{\sout{#1} \textcolor{red}{#2}}
\newcommand{\add}[1]{\textcolor{red}{#1}}

\renewcommand{\delete}[1]{}
\renewcommand{\replace}[2]{\textcolor{black}{#2}}
\renewcommand{\add}[1]{\textcolor{black} {#1}}

\newcommand{\adds}[1]{\textcolor{black}{#1}}

\usepackage{hyperref}

\begin{document}



\title{
Strong hole-photon coupling in planar Ge for probing charge degree and strongly-correlated states
}


\author[1,2]{Franco De Palma}
\equalcont{These authors contributed equally to this work.}

\author[1,2]{Fabian Oppliger}
\equalcont{These authors contributed equally to this work.}

\author[1,2]{Wonjin Jang}
\equalcont{These authors contributed equally to this work.}

\author[3,4]{Stefano Bosco}
\author[5]{Marián Janík}
\author[6]{Stefano Calcaterra}
\author[5]{Georgios Katsaros}
\author[6]{Giovanni Isella}
\author[3]{Daniel Loss}
\author*[1,2]{Pasquale Scarlino}\email{pasquale.scarlino@epfl.ch}

\affil[1]{\orgdiv{Hybrid Quantum Circuit Laboratory, Institute of Physics and Center for Quantum Science and Engineering}, \orgname{École Polytéchnique Fédérale de Lausanne (EPFL)}, \orgaddress{\city{Lausanne}, \postcode{1015}, \country{Switzerland}}}
\affil[2]{\orgdiv{Center for Quantum Science and Engineering}, \orgname{École Polytéchnique Fédérale de Lausanne (EPFL)}, \orgaddress{\city{Lausanne}, \postcode{1015}, \country{Switzerland}}}

\affil[3]{\orgdiv{Department of Physics}, \orgname{University of Basel}, \orgaddress{\street{Klingelbergstrasse 82}, \city{Basel}, \postcode{4056}, \country{Switzerland}}}

\affil[4]{\orgdiv{QuTech}, \orgname{Delft University of Technology}, \orgaddress{\city{Delft}, \country{The Netherlands}}}

\affil[5]{\orgname{Institute of Science and Technology Austria}, \orgaddress{\street{Am Campus 1}, \city{Klosterneuburg}, \postcode{3400}, \country{Austria}}}

\affil[6]{\orgdiv{L-NESS, Physics Department}, \orgname{Politecnico di Milano}, \orgaddress{\street{via Anzani 42}, \city{Como}, \postcode{22100}, \country{Italy}}}

\abstract{
Semiconductor quantum dots (QDs) in planar germanium (Ge) heterostructures have emerged as front-runners for future hole-based quantum processors. 
Here, we present strong coupling between a hole charge qubit, defined in a double quantum dot (DQD) in planar Ge, and microwave photons in a high-impedance ($Z_\mathrm{r} = 1.3 ~\mathrm{k}\Omega$) resonator based on an array of superconducting quantum interference devices (SQUIDs). 
Our investigation reveals vacuum-Rabi splittings with coupling strengths up to $g_0/2\pi = 260 ~\mathrm{MHz}$, and a cooperativity of $C \sim 100$, dependent on DQD tuning.
Furthermore, utilizing the frequency tunability of our resonator, we explore the quenched energy splitting associated with \replace{strongly correlated Wigner molecule (WM) states that emerge}{strong Coulomb correlation effects} in Ge QDs. 
The observed enhanced coherence of the \replace{WM}{strongly correlated} excited state signals the presence of distinct symmetries within related spin functions, serving as a precursor to the strong coupling between photons and spin-charge hybrid qubits in planar Ge.
This work paves the way towards coherent quantum connections between remote hole qubits in planar Ge, 
required to scale up hole-based quantum processors.
}


\maketitle

\newpage
\section{Introduction}\label{sec1}

Semiconductor quantum dots (QDs) represent a promising platform for advanced quantum information processing \cite{2023_bur, 2022_tak, 2022_phi}. Particularly, hole confinement in QDs enables rapid electric spin manipulation due to the large spin-orbit interaction \cite{2020_hen, 2021_fro, 2021_sca, 2007_bul, 2005_bul}.
QD-based hole qubit systems have been implemented in various platforms including fin field-effect transistors (finFETs) 
\cite{2016_voi, 2022_gey},
Ge/Si core/shell nanowires \cite{2011_klo, 2021_fro}, and planar Ge/SiGe heterostructures \cite{2021_hen, 2020_hen, 2020_hena}.
Among these, planar Ge stands out due to its exceptional characteristics \cite{2021_sca}, including high hole mobility ($\mu $ $>$ $10^6 ~\mathrm{cm^2V^{-1}s^{-1}}$ \cite{2023_ste}), low charge noise \cite{Nico_first}, and a low effective mass \cite{Ge_meff}.
Furthermore, nuclear isotope purification can be performed, effectively mitigating magnetic field noise and enhancing the qubit coherence \cite{2021_sca}. 
Building on all these advantages, recent works have demonstrated coherent single- and two-qubit operations \cite{2020_hen, 2020_hena}, scalable multi-qubit array architecture \cite{2021_hen, 2023_bor}, and coherent spin shuttling \cite{2023_van} in planar Ge.

In the context of circuit quantum electrodynamics (cQED), the hybridization of microwave photons in superconducting cavities with QD-based qubits holds enormous potential for various applications in quantum technology. 
These applications include enabling long-range interactions between distant quantum dot qubits \cite{2022_har, 2018_van, 2020_bor}, achieving rapid and high-fidelity charge and spin state detection \cite{2019_zhe, 2019_sca}, as well as facilitating analog quantum simulation of open quantum systems \cite{2022_kim}, and advancing the development of gigahertz photodetectors \cite{2021_kha}.
However, achieving strong light-matter coupling is a fundamental prerequisite for these endeavors. 
While several previous experiments have successfully demonstrated strong coupling for electrons hosted in Si \cite{2017_mi, 2018_sam, 2018_mi}, GaAs \cite{2022_sca, 2017_sto}, 
InAs nanowires \cite{2023_ung}, and for holes in silicon \replace{finFETs}{nanowire transistors} \cite{2023_Grenoble}, the strong coupling of holes in planar Ge has remained elusive \cite{2023_kan, 2019_wan, 2018_li}.

Previous hybrid cQED experiments primarily focused on resonators interacting with the ground and first excited states of double quantum dot (DQD) charge- or spin-two level systems. 
In fact, in typical QD structures, additional single-dot orbital states usually lie at energies higher than $100 ~ h \cdot$GHz, making them inaccessible to microwave resonators \cite{2003_han}.
However, low excitation energies can arise from Coulomb interaction-induced \replace{redistribution of multiple charges confined}{renormalization of orbital energies} in \delete{a} single QD\add{s},
leading to the formation of strongly correlated states \add{(SCSs)} \cite{2008_gol, 2007_yan}.
\replace{These states, referred to as Wigner molecules (WM) \cite{2007_yan} and facilitated by anisotropic QD confinement \cite{2021_aba},}
{When further enhanced by anisotropic QD confinement \cite{2021_aba}, these states can lead to excitation energies below $10 ~ h \cdot$GHz that have been observed in GaAs \cite{2023_jan, 2021_jan}, Si \cite{2021_cor}, and carbon nanotube \cite{2013_pec, 2019_sha} QDs \adds{and attributed to Wigner molecular states \cite{2022_yana, Erc_2021, 2022_yan, 2007_yan, 1999yannouleas, 2009li, 2004szafran}}.
The emergence of SCSs is a general phenomenon, which can take place in QDs defined in any semiconductor platform \cite{2013_pec, 2021_jan, 2021_cor}.
Such SCSs have profound implications for quantum information processing, offering an encoding for spin-charge hybrid qubits based on exchange interaction \cite{2022_yan, 2021_jan}.}
If not properly controlled, it can significantly reduce the fidelity of conventional readout schemes in spin qubits \cite{2021_aba}.
In Ge, it has been also shown that SCSs enable anomalous splittings of spin energy levels without the need for magnetic fields \cite{2022_het}. 
These findings suggest that \replace{WMs}{low-lying SCSs} could serve as a valuable interface between QD qubits and superconducting circuits in hybrid architectures.

In this study, we establish strong coupling between a microwave photon and a DQD-based hole charge qubit in a planar Ge/SiGe heterostructure, using a high-impedance frequency-tunable resonator based on superconducting quantum interference devices (SQUIDs) \cite{2017_sto}. 
We explore different DQD configurations and achieve a charge-photon vacuum-Rabi splitting (charge decoherence rate) up to $2g_0/2\pi\sim 520$~MHz (down to $\Gamma/2\pi \sim 57$~MHz). We estimate a system cooperativity of $C \sim 100$, among the highest reported for QDs charge-resonator hybrid systems to date \cite{2022_sca}. 
Our device geometry facilitates formation of \replace{WMs}{SCSs} in Ge, unveiling a quenched energy spectrum of \replace{WM states}{SCSs} in the DQD. 
Leveraging the frequency tunability of the SQUID array resonator, we perform resonant energy spectroscopy of \replace{WM states}{SCSs} in the DQD and extract their energy spectra.
By exploring several pairs of adjacent inter-dot configurations, we observe selective coupling to the resonator based on the parity of the DQD hole number and enhanced coherence times for certain excited \replace{WM states}{SCSs}, which we attribute 
to states with a different spin structure \cite{2012_shi, 2021_jan}.

\newpage
\section{Results}

\subsection*{Architecture for hybrid circuit QED with holes in planar Ge}

\begin{figure}[H]

     \centering
     \includegraphics[width=1\textwidth]{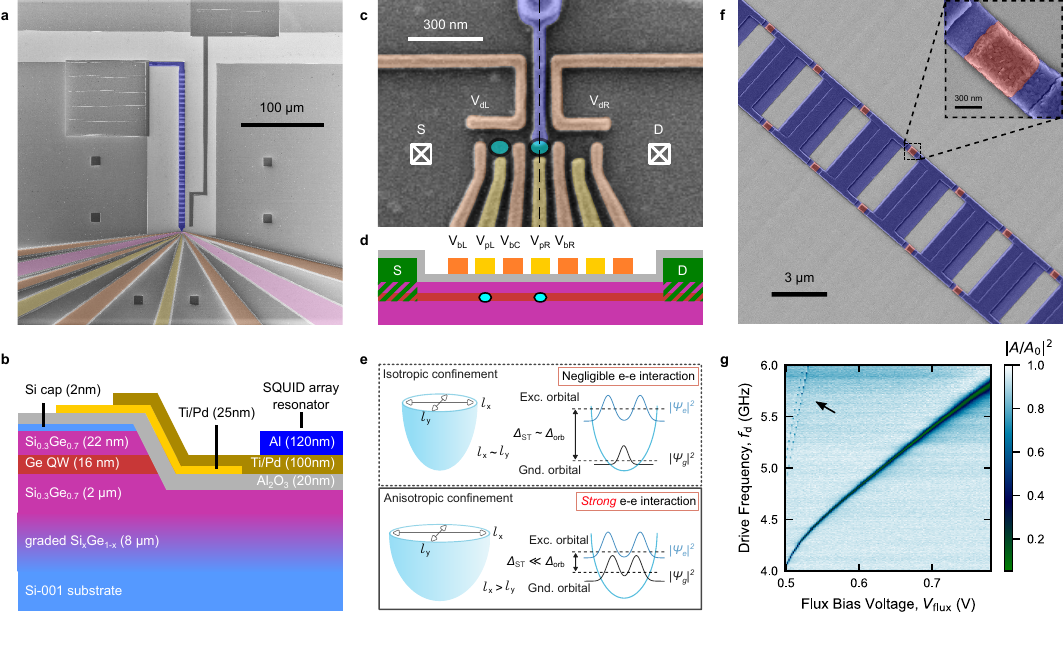}
     \caption{
     \textbf{Superconductor-semiconductor hybrid architecture on planar Ge heterostructure.} 
     \textbf{a,} False-colored scanning electron micrograph of a representative hybrid device.
     The SQUID array resonator (violet) is capacitively coupled to the transmission line on top.
     The QDs are defined electrostatically by barrier (orange) and plunger (yellow) gates.
     The Ge quantum well is etched away everywhere except for a small mesa region (pink) used to host the QDs.
     Ohmic contacts are patterned on the extensions of the mesa region. 
     \textbf{b,} Schematic side-view of the heterostructure and the device across the black dashed line in \textbf{c}. 
     \textbf{c,} False-colored scanning electron micrograph of the QDs region. The expected position of the DQD is highlighted by cyan ellipses. 
     The plunger gate $V_\text{pL}$ ($V_\text{pR}$) mainly controls the electrochemical potential of the left (right) QD, while $V_\text{bL}$ ($V_\text{bR}$) modulates the tunnel coupling strength of the left (right) QD to left (right) reservoir.
     $V_\text{bC}$ controls the inter-dot tunnel coupling strength $t_\text{c}$.
     \textbf{d,} Side-view of the device across the QD array. 
     \textbf{e,} Schematic of the \add{two-body} ground and excited state wavefunctions ($\psi_\text{g}$ and $\psi_\text{e}$) and single QD energy splitting for two different classes of QD confinement potential. Under isotropic confinement, the ground and excited state wavefunctions 
     have distinct shapes, which result in large orbital splitting $\Delta_\text{orb}$. 
     In the anisotropic and strongly interacting case, the symmetry of $\psi_\text{g}$ is broken, resulting in a quenched singlet-triplet splitting $\Delta_\mathrm{ST} \ll \Delta_\text{orb}$ \cite{2021_aba}.
     \textbf{f,} False-colored scanning electron micrograph of the SQUID array resonator. Inset: Zoom-in of a single Josephson junction (red). 
     \textbf{g,} Flux tunability of the SQUID array resonator.
     Normalized amplitude of feedline transmission $|A/A_\mathrm{0}|^2$ as a function of drive frequency $f_\mathrm{d}$ and bias voltage $V_\text{flux}$ applied to the superconducting coil mounted perpendicularly to the sample (see Supplementary Note~\ref{sup:supplsec1}).
    Higher resonator modes are visible near the half-flux point (black arrow) \cite{2012_Masluk}.
     The device is operated in a dilution refrigerator with a base temperature of 10 mK
     (see Supplementary Note~\ref{sup:supplsec1}).
     }
     \label{fig:Fig1}
\end{figure}

Fig.~\ref{fig:Fig1}a shows the hybrid superconductor-semiconductor device fabricated on a Ge/SiGe heterostructure \cite{2021_jir}.
As shown in Fig.~\ref{fig:Fig1}b, the 16 nm Ge quantum well (QW), hosting the 2-dimensional hole gas (2DHG), is $\sim$~24~nm below the surface.
A conductive channel, defined by selectively etching the Ge QW, hosts a DQD (cyan ellipses in Figs.~\ref{fig:Fig1}c, d) defined by metallic gate electrodes. 
The gate layout of our device supports relatively large QDs (\replace{diameter $\sim$ 100}{radius $l_\mathrm{QD} \sim 70$}~nm).
\replace{Since the Coulomb interactions significantly exceed the single-particle orbital energy resulting from confinement, we anticipate a decreased singlet-triplet energy gap $\Delta_\mathrm{ST}$ \cite{2008_gol}, as illustrated in Fig.~\ref{fig:Fig1}e.}{
The Wigner ratio $\lambda_\mathrm{W} = E_\mathrm{ee} / E_\mathrm{orb} \propto l_\mathrm{QD}$ (see Supplementary Note~\ref{sup:wigner}) quantifies the ratio between the Coulomb interaction strength ($E_\mathrm{ee} \propto 1/l_\mathrm{QD}$) and the orbital confinement energy ($E_\mathrm{orb} \propto 1/l_\mathrm{QD}^2$).
Coulomb interactions become increasingly relevant in large QDs, as the ones studied here. 
In our experiment, we estimate $\lambda_\mathrm{W} \sim 4.46$.
Coulomb correlation renormalizes the energy of orbital states in QDs, thus quenching the orbital splitting and therefore the singlet-triplet splitting $\Delta_\mathrm{ST}$ \cite{2008_gol}.
}
\replace{This Coulomb interaction fosters the emergence of Wigner molecule (WM) states, even in the presence of a slight anisotropy in the QD (see Supplementary Note~\ref{sup:io_cq}) \cite{2021_aba}}
{Furthermore, anisotropic QD confinement is expected to enhance the correlation effect and reduce $\Delta_\mathrm{ST}$ even further (see Supplementary Note~\ref{sup:wigner}) \cite{2021_aba}, as illustrated in Fig.~\ref{fig:Fig1}e. 
Orbital state renormalization induced by Coulomb correlation and confinement anisotropy is expected to significantly alter also the charge density distribution of the ground state, promoting the formation of Wigner molecular (WM) states (see Supplementary Figure~\ref{sup:WM_Levels}) \cite{2021_aba, 2023_jan, 2021_cor, 2013_pec, 2019_sha}.
}

The right dot is coupled to the superconducting resonator (Fig.~\ref{fig:Fig1}f) via the violet electrode in Fig.~\ref{fig:Fig1}c (see Supplementary Note~\ref{sup:supplsec1}) \cite{2017_sto}. This is designed 
to maximize the capacitive coupling by completely overlapping one QD and, therefore, to efficiently couple to the DQD via transverse charge-photon interaction
\cite{2017_sto, 2022_sca, 2017_mi}.
The resonator consists of an array of $N = 32$ SQUIDs (Fig.~\ref{fig:Fig1}f) with an inductance of $L \sim 0.63$ nH/SQUID, resulting in an equivalent lumped impedance of $Z_\mathrm{r} \sim 1.3 ~\text{k}\Omega$
\cite{2017_sto}.
The high-impedance resonator enhances the \replace{qubit}{charge}-photon coupling strength $g_0$ by maximizing the vacuum voltage fluctuation $V_\mathrm{0,rms}=2\pi f_\mathrm{r}\sqrt{\hbar Z_\mathrm{r}/2} $, 
according to the relation $g_\mathrm{0}=\frac{1}{2}\beta_\mathrm{r} V_\mathrm{0,rms}/\hbar$, with $\beta_\mathrm{r}$ the resonator differential lever arm 
\cite{2023_Grenoble}.
The resonator is also capacitively coupled to a 50 $\Omega$ waveguide (the photon feedline) on one side, and grounded on the other end, forming a hanged quarter-wave resonator (Fig.~\ref{fig:Fig1}a) \cite{2017_sto}.
We probe the microwave response of the hybrid system recording the feedline transmission ($S_\text{21}$) at powers corresponding to less than one photon on average in the resonator (see Supplementary Note~\ref{sup:cav_char}).
By leveraging the external magnetic flux dependence 
of the critical current of the SQUIDs \cite{JPA}\delete{ (see Supplementary Note~\ref{sup:supplsec1})}, the resonator frequency $f_\text{r}$ can be tuned from $\sim 6$~GHz to well below 4~GHz (see Fig.~\ref{fig:Fig1}g).
\add{To apply a finite magnetic flux, we place a superconducting coil on top of the device which generates an out-of-plane magnetic field of $50 \sim 70\ \mu$T (see Supplementary Note~\ref{sup:supplsec1}).}

\begin{figure}[H]
     \centering
     \includegraphics[width=95mm]{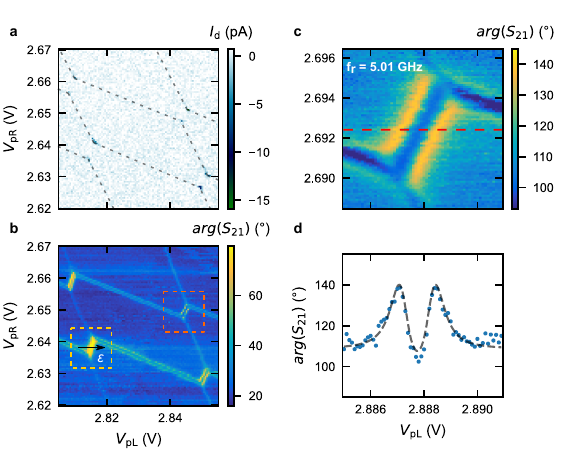}
     \caption{ 
     \textbf{DQD characterization with the tunable resonator.} 
     A region of the DQD charge stability diagram as a function of the applied plunger gate voltages $V_\text{pR}$ and $V_\text{pL}$, recorded by dc-transport (\textbf{a}) and by measuring \replace{amplitude (\textbf{b}) and }{the} phase (\textbf{b}) of the feedline transmission $S_{21}$, $arg(S_{21})$, at $f_\text{d} = f_\text{r} = 5.01$~GHz. The resonator detects inter-dot and reservoir-dot transitions when 
     their tunneling rates are close to $f_\text{r}$ \cite{2012_fre}.
     Yellow (orange) dashed box in \textbf{b}: 
     The phase signal increases (decreases) near the inter-dot region with respect to the background, if the resonator is dispersively shifted to lower (higher) frequency. Notably, because the resonator gate lever arm is larger for the right QD, the resonator is more sensitive to its QD-reservoir transitions with respect to those of the left QD.
     \textbf{c,} \delete{Same }Inter-dot transition probed with \delete{$f_\text{r} = 4.5 ~\mathrm{GHz} < 2t_\text{c}/h$ (\textbf{d}), }$f_\text{r} = 5.01 ~\mathrm{GHz} \sim 2t_\text{c}/h$.
     \delete{(\textbf{e}) and $f_\text{r} = 5.5 ~\mathrm{GHz} > 2t_\text{c}/h$ (\textbf{f})}.
     \replace{The corresponding line-cuts, taken along the red dashed lines, are shown in \textbf{g}, \textbf{h}, and \textbf{i} respectively.}{\textbf{d,} A line-cut taken along the red dashed line in \textbf{c}.}
     The black dashed curve shows the \delete{simultaneous }fit to a master equation (see Methods).
     }
     \label{fig:Fig2}
\end{figure}

Fig.~\ref{fig:Fig2}a shows a region of the DQD stability diagram spanned by $V_\text{pR}$ and $V_\text{pL}$, measured by probing the dc current through the DQD \cite{2003_DQD_rev}.
To characterize the charge-photon coupling, we simultaneously monitor the feedline transmission at the frequency $f_\text{d} = f_\text{r} = 5.01$~GHz (see Fig.~\ref{fig:Fig2}b).
While the dc transport measurement for the explored configuration only exhibits the DQD triple points
\cite{2003_DQD_rev}, the resonator response reveals not only the inter-dot transitions, but also the QD-reservoir ones,
facilitating an extensive characterization of QD devices.
\replace{Extended Data Figure~1}{Supplementary Figure~\ref{sup:Extended data 2}} reports a zoom-out of the charge stability diagram shown in Fig.~\ref{fig:Fig2}a and b.

Close to an inter-dot transition, the DQD system can be approximated by a simplified $2 \times 2$ charge qubit Hamiltonian,
given by $H_\text{cq} = \frac{\varepsilon}{2} \sigma_z + t_\text{c} \sigma_x$ with corresponding eigenenergies $E_\pm = \pm \frac{1}{2}\sqrt{\varepsilon^{2} + 4t_\text{c}^2}$. Here, $\sigma_i$ represents the Pauli operator $(i = x, y, z)$ \cite{2017_sto} and $\varepsilon$ ($t_\mathrm{c}$) is the DQD energy detuning (tunnel coupling). 
The transverse charge-photon interaction $H_\text{int} = \hbar g_\text{eff} (a^\dag \sigma^- + a \sigma^+)$, with $g_\mathrm{eff}=2g_\mathrm{0}t_\mathrm{c}/(E_+-E_-)$ denoting the effective charge-photon coupling strength, hybridizes the qubit with the resonator (see Supplementary Note~\ref{sup:io_cq}).
As a result, the phase of the feedline transmission $S_{21}$ (Fig.~\ref{fig:Fig2}b) exhibits a different response depending on whether the qubit energy is higher (yellow dashed box) or lower (orange dashed box) than the bare resonator energy.

While tuning the qubit frequency, $f_\mathrm{q}=(E_+-E_-)/h = \sqrt{\varepsilon^{2} + 4t_\text{c}^2}/h$, to be close to $f_\mathrm{r}$ ($|f_\mathrm{q} - f_\mathrm{r}| < 10 g_0/2\pi$) is essential to ensure a significant dispersive resonator response,
it can be challenging to achieve depending on the DQD gate layout \cite{2023_Grenoble}.
The tunable resonator presented here offers an additional means to efficiently investigate qubits by varying $f_\mathrm{r}$ across $f_\mathrm{q}$. 
In Fig.~\ref{fig:Fig2}c,
we record the phase of the feedline transmission\add{, arg($S_{21}$), taken with a resonator frequency of $f_\mathrm{r} = 5.01~\mathrm{GHz} \sim 2t_\mathrm{c}/h$ to reconstruct the DQD stability diagram of an inter-dot transition.}
\delete{for the same inter-dot transition at three different values of $f_\text{r}$: lower than (Fig.~\ref{fig:Fig2}d), resonant with (Fig.~\ref{fig:Fig2}c) and higher than (Fig.~\ref{fig:Fig2}f) $2t_\text{c}/h \sim 5$~GHz.}
\add{The corresponding line-cut along the red dashed line is reported in Fig.~\ref{fig:Fig2}d.
Leveraging the frequency tunability of our resonator, we also measure the same region of the DQD stability diagram with $f_\mathrm{r}$ tuned above and below $2t_\text{c}/h$, reported in Supplementary Figure~\ref{sup:Fig2}.}
Line-cuts across the inter-dot transition (\replace{Fig.~\ref{fig:Fig2}g - i}{Fig.~\ref{fig:Fig2}d and Supplementary Figures~\ref{sup:Fig2}g - i}) are simultaneously fitted to a master equation model (\add{denoted by }black dashed lines\delete{in Fig.~\ref{fig:Fig2}g-i}), extracting a common tunnel coupling of $t_\text{c}/h = 2.472$~GHz, qubit decoherence $\Gamma/2\pi = 120$~MHz, and charge-photon coupling strength\delete{s} \replace{$(g_0^\mathrm{(g)}/2\pi, g_0^\mathrm{(h)}/2\pi, g_0^\mathrm{(i)}/2\pi) = (162, 192, 172)$~MHz}{$g_0/2\pi = 192 $~MHz} for $f_\mathrm{r} = 5.01~\mathrm{GHz} \sim 2t_\mathrm{c}/h$ (see Methods).
To quantify the quality of the DQD-resonator interface, we evaluate the ratio between the coupling and the decoherence rates, by computing the cooperativity $C = {4g_0^2}/({\kappa\Gamma})$ 
\cite{Coop_Clerk}.
Using $\kappa/2\pi = 30$~MHz, extracted from a bare resonator fit at 5~GHz, along with the aforementioned parameters, we estimate $C \sim 40\gg1$, indicating the possibility to observe strong coupling.

\subsection*{Strong hole charge-photon coupling}

\begin{figure}[H]
     \centering
     \includegraphics[width=95mm]{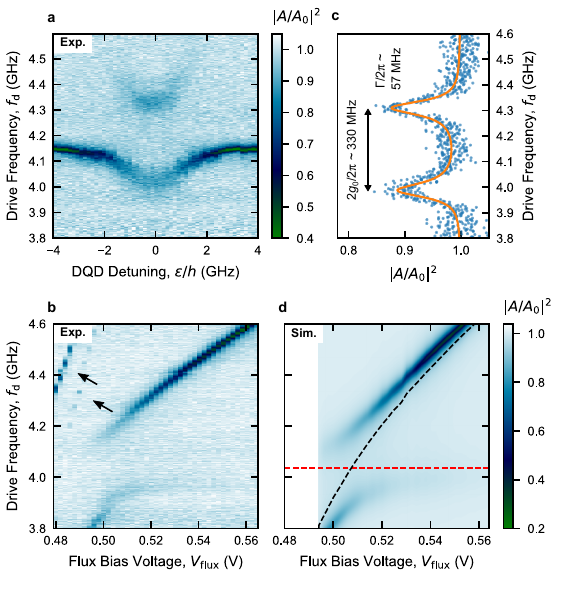}
     \caption{\textbf{Strong charge-photon coupling at the charge sweet spot.}
     \textbf{a,} Normalized amplitude of feedline transmission $|A/A_\mathrm{0}|^2$ as a function of drive frequency $f_\mathrm{d}$ and DQD detuning $\varepsilon$. An avoided crossing - the signature of the strong coupling regime - is observed when the DQD-charge transition matches the bare resonator frequency.
     \textbf{b,} $|A/A_\mathrm{0}|^2$ as a function of drive frequency $f_\mathrm{d}$ and the voltage $V_\text{flux}$ applied to the resonator coil, which tunes the resonator frequency $f_\text{r}$. During the measurement, the DQD is kept at $\varepsilon = 0$.
     An avoided crossing is observed around $V_\text{flux} = 504$ mV, when the bare resonator frequency $f_\text{r}$ matches the DQD charge transition ($f_r=f_q=2t_\text{c}/h$).
     Higher resonator modes are visible near the half-flux point (black arrows) \cite{2012_Masluk}.
     \delete{\textbf{e,} Frequency line-cut (along the orange dashed line in \textbf{d} at the avoided crossing.
     A fit to the master equation model is represented by a solid orange line (see Methods). The extracted values for $2g_0$ and $\Gamma$ are indicated (2$t_\text{c}/h = 4.036$ GHz).}
     \textbf{c,} \replace{Frequency line-cut at resonance}{$|A/A_\mathrm{0}|^2$ as a function of $f_\mathrm{d}$ at the resonance condition}\delete{ (along the orange dashed line in \textbf{a})}, highlighting the vacuum-Rabi splitting $2g_0/2\pi$.
     A fit to the master equation model is represented by a solid orange line (see Methods).
     All the extracted values are reported in Supplementary Table 1. $2g_0$ and $\Gamma$ are indicated (2$t_\text{c}/h = 4.149$~GHz).
     \textbf{d,} Simulation of $|A/A_0|^2$ using input-output theory with the parameters \replace{extracted
     from the fit in panel \textbf{e} (see Methods).}{$g_0/2\pi=154$~MHz, $\Gamma/2\pi=80$~MHz, $t_\text{c}/h = 2.018$~GHz extracted from fitting a line-cut of panel \textbf{b} at $V_\text{flux} = 504$ mV (reported in Supplementary Figure~\ref{sup:fig3_flux_sweep}b).
     }
     }
     \label{fig:Fig3}
\end{figure}

We now probe the charge-photon interaction at $\varepsilon=0$ (\add{charge} sweet spot), where the electric dipole moment of the holes in the DQD is maximal, \add{resulting in a vacuum-Rabi mode splitting of $2g_\mathrm{eff} =2g_0$ \cite{2017_mi, 2017_sto}.
}
Fig.~\ref{fig:Fig3}a shows the normalized feedline transmission amplitude $|A/A_0|^2$ as a function of $f_\text{d}$ and with $\varepsilon$ changed to cross an inter-dot transition (as depicted by the \replace{white}{black} arrow in Fig.~\ref{fig:Fig2}b).
\replace{
Near $\varepsilon=0$, the two subsystems maximally hybridize, resulting in a vacuum-Rabi mode splitting of $2g_\mathrm{eff} = 2g_0$ \mbox{\cite{2017_mi,2017_sto}},
represented by the line-cut in Fig.~\ref{fig:Fig3}b along the orange dashed line in Fig.\ref{fig:Fig3}a.
}{
We note that, as we detail in Fig.~\ref{fig:Fig4} below, the two subsystems are not perfectly in resonance at $\varepsilon=0$ in Fig.~\ref{fig:Fig3}a.}
Our resonator's frequency tunability offers a convenient way to investigate vacuum-Rabi splitting while keeping the DQD electrostatic configuration constant. 
This allows us to reach the resonant condition between the DQD two-level system and the resonator, while keeping the DQD gate voltages unchanged. 
\add{
Thereby, we fix the detuning at $\varepsilon = 0$ and vary the external magnetic flux to fine-tune the resonator frequency $f_\mathrm{r}$ into resonance with the qubit frequency $f_\mathrm{q}=2t_\mathrm{c}/h$.
In Fig.~\ref{fig:Fig3}b, we report $|A/A_0|^2$ as a function of $f_d$ and flux bias voltage $V_\mathrm{flux}$, where the charge-photon hybridization at $\sim 4$~GHz results in a clear vacuum-Rabi mode splitting.
By fitting a line-cut of Fig.~\ref{fig:Fig3}b taken at $V_\mathrm{flux} = 504$~mV (reported in Supplementary Figure~\ref{sup:fig3_flux_sweep}b), we extract the parameters $(t_\mathrm{c}/h, g_0/2\pi, \Gamma/2\pi) = (2018, 154, 80)$~MHz.
These parameters are utilized to numerically reconstruct $|A/A_0|^2$ (Fig.~\ref{fig:Fig3}d). To better evaluate the cooperativity of our system, in Fig.~\ref{fig:Fig3}c we report a high-quality vacuum-Rabi mode splitting measured, with increased averaging, as a function of $f_\mathrm{d}$ in the same DQD configuration ($\varepsilon = 0$), but at $V_\mathrm{flux} = 507$~mV to compensate for a slight drift in qubit frequency.
}
By fitting to the master equation model (solid line in Fig.~\ref{fig:Fig3}c, see Methods), we find
$(t_\mathrm{c}/h, g_0/2\pi, \Gamma/2\pi) = (2072, 165, 57)$~MHz.
\delete{The simulated feedline transmission, based on these parameters, closely matches the experimental data (see Fig.~\ref{fig:Fig3}c).}
These parameters result in a cooperativity of $C \sim 100$ (with $\kappa/2\pi = 19$~MHz), which highlights the strong charge-photon coupling in planar Ge.

In \replace{Extended Data Figure 2a}{Supplementary Figure~\ref{sup:ext_fig2}a}, we explore an alternative DQD charge transition, which features an enhanced $g_0$.
Fitting the line-cut in \replace{Extended Data Figure 2b}{Supplementary Figure~\ref{sup:ext_fig2}b} to the master equation model, we extract the parameters $(t_\mathrm{c}/h, g_0/2\pi, \Gamma/2\pi) = (2711, 260, 192)$~MHz,
and calculate a cooperativity of $C \sim 23$ (with $\kappa/2\pi = 63$~MHz).
Here, the high $g_0/2\pi = 260$~MHz, enabled by the high-impedance SQUID array resonator, allows us to achieve the strong coupling regime, in spite of a substantial qubit decoherence rate $\Gamma$.
We speculate that the difference between the values of $g_{0}$ and $\Gamma$ extracted from the two datasets in Fig.~\ref{fig:Fig3} and \replace{Extended Data Figure 2}{Supplementary Figure~\ref{sup:ext_fig2}} may arise from distinct effective electric dipole moments associated with the two DQD electrostatic configurations \cite{2022_sca}. 
To account for the frequency dependence of the coupling strength between the resonator and DQD, we calculate the resonator's differential lever arm $\beta_\mathrm{r}$ = $\frac{2g_\mathrm{0}\hbar}{V_\mathrm{0,rms}}$ in the two configurations. We find $\beta_\mathrm{r}$ values of 0.18 and 0.25 eV/V (see Methods),
respectively, indicating a higher coupling of the resonator to the detuning degree of freedom in the second case, albeit at the cost of a larger $\Gamma$ \cite{2022_sca}.

\delete{
Our resonator's frequency tunability offers a convenient way to investigate vacuum-Rabi splitting while keeping the DQD electrostatic configuration constant. 
This allows us to reach the resonant condition between the DQD two-level system and the resonator, while keeping the DQD gate voltages unchanged. 
The line-cut at the resonant point, taken along the orange dashed line in Fig.~\ref{fig:Fig3}d, reported in Fig.~\ref{fig:Fig3}e, reveals vacuum-Rabi mode splitting of the DQD-resonator hybridized system.
Fitting the line-cut to the master equation model, we find the parameters $(t_\mathrm{c}/h, g_0/2\pi, \Gamma/2\pi) = (2018, 154, 80)$~MHz, corresponding to a cooperativity $C\sim52$ (with $\kappa/2\pi = 23$~MHz).
The discrepancy between the parameters extracted by sweeping $\varepsilon$ (Figs.~\ref{fig:Fig3}a and b) and $V_\mathrm{flux}$ (Figs.~\ref{fig:Fig3}d and e)
is attributed to a jump of the charge qubit to a slightly lower $t_\mathrm{c}$ and to the shorter integration time used for the latter measurement.
The extracted system parameters obtained from the different measurements are summarized in Extended Data Table 1.
}

\subsection*{Tunable high-impedance resonator for qubit spectroscopy}
\begin{figure}[H]
     \centering
     \includegraphics[width=95mm]{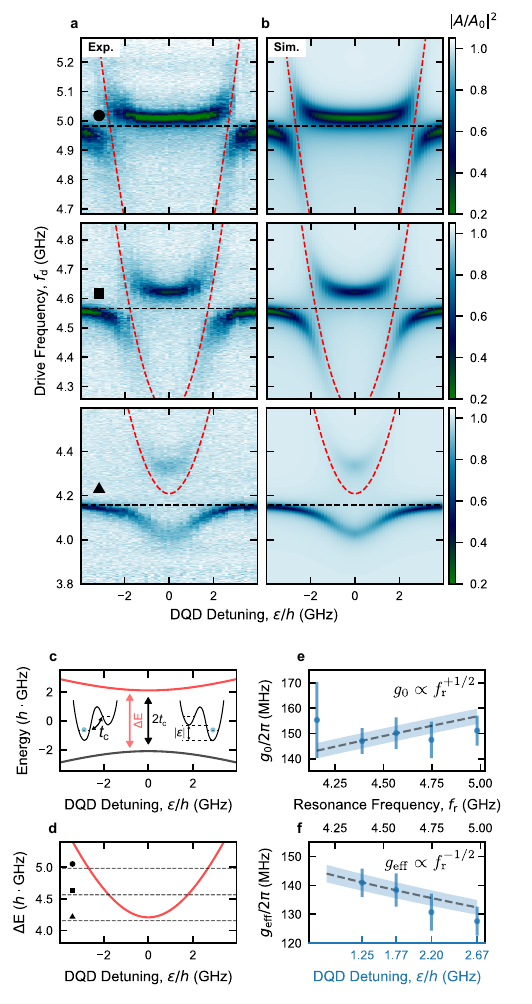}
     \caption{\textbf{Charge qubit spectroscopy via tunable resonator.} 
     \textbf{a,} Normalized amplitude of feedline transmission $|A/A_\mathrm{0}|^2$ as a function of drive frequency $f_\mathrm{d}$ and DQD detuning $\varepsilon$. The three panels are taken in correspondence of three different bare resonator frequencies $f_\text{r}$, denoted by black circle, square and triangle, while keeping the inter-dot tunnel coupling $t_\text{c}$ constant.
     \textbf{b,} Simulation of $|A/A_\mathrm{0}|^2$
     using input-output theory with the parameters extracted 
     by fitting the full dataset in the corresponding panels in \textbf{a} to the master equation model (see Methods).
     Black (red) dashed lines in \textbf{a} and \textbf{b} denote the bare resonator (DQD-charge qubit) frequency.
     \textbf{c,} Energy-level diagram, i.e. the energy spectrum, of the DQD charge qubit system \delete{(top panel) and the excitation energy $\Delta E$ (bottom panel) }as a function of DQD detuning $\varepsilon$ (calculated for the charge qubit Hamiltonian in main text).
     \replace{Top panel:}{The} black (red) curve represents the ground (excited) state of the charge qubit.
     Inset: DQD potential schematics showing the charge state at the negative and positive $\varepsilon$. 
     \add{\textbf{d,} Excitation energy $\Delta E$ as a function of DQD detuning $\varepsilon$.}
     \delete{Bottom panel:}
     The dashed lines denoted by black circle, square and triangle correspond to different $f_\text{r}$ in \textbf{a}.
     \textbf{e,} Extracted \add{charge-photon }coupling strength \add{$g_0$} as a function of $f_\mathrm{r}$.
     \replace{Charge-photon coupling strength $g_0$ (blue dots) does not show an explicit dependence on $f_\mathrm{r}$, whereas $g_\text{eff}$ decreases along $f_\mathrm{r}$ due to the decreasing dipole moment of the charge qubit.
     The black dashed line shows the predicted dependence of $g_\mathrm{eff}=g_02t_\mathrm{c}/hf_\text{r}$.
     }{
     \textbf{f,} Effective charge-photon coupling strength $g_\mathrm{eff}$ at the $\varepsilon$ values for which the two subsystems are in resonance, estimated using $g_\mathrm{eff}=g_02t_\mathrm{c}/hf_\text{r}$. The DQD detuning values corresponding to the avoided crossings are indicated at the bottom axis in blue.
     Since the resonance condition is not met for the lowest panel in \textbf{a} ($f_\mathrm{r} < 2t_\mathrm{c}/h$), the first point in \textbf{f} is omitted.
     The dashed lines in \textbf{e} and \textbf{f} represent the expected trend of the coupling strengths as a function of $f_\mathrm{r}$,
     while the error bars and the shaded regions indicate the estimated uncertainties (see Methods).
     }
     }
     \label{fig:Fig4}
\end{figure}

We leverage the resonator frequency tunability to conduct resonant energy spectroscopy of the DQD charge qubit, in the same DQD configuration as in Fig.~\ref{fig:Fig3}, and keeping the DQD at a fixed $t_\mathrm{c}$ \cite{2017_sto}. 
This spectroscopy aims to reconstruct the qubit's energy dispersion. 
In contrast to the measurements in Fig.~\ref{fig:Fig3}, where $f_\mathrm{r} \sim 2 t_\mathrm{c}/h$, here we extend our investigation also to higher resonator frequencies, $f_\mathrm{r} > 2 t_\mathrm{c}/h$.

In Fig.~\ref{fig:Fig4}a, we present the measured normalized feedline transmission $|A/A_0|^2$ as a function of $f_\text{d}$ and $\varepsilon$ for three different values of $f_\text{r}$, as indicated by dashed lines in \delete{the bottom panel of }Fig.~\ref{fig:Fig4}d. 
The schematics in Figs.~\ref{fig:Fig4}c \add{and d} illustrate the charge qubit energy level diagram (\replace{top}{panel~c}) and the excitation energy spectrum $\Delta E= E_+ - E_-= E_\text{cq}$ (\replace{bottom}{panel~d}) along $\varepsilon$. 
Notably, clear avoided crossings are observed in Fig.~\ref{fig:Fig4}a when the charge qubit gets in resonance with the resonator ($h f_\mathrm{r}=\sqrt{\varepsilon^2 + 4t_\mathrm{c}^2}$). 
Additional details on the charge qubit spectroscopy are available in \replace{Extended Data Figure 3}{Supplementary Note~\ref{sup:cq_spectrum}}.

These spectroscopy measurements also provide valuable insights into the evolution of the effective charge-photon coupling strength $g_\mathrm{eff}$, as a function of the DQD detuning $\varepsilon$. 
By fitting all five datasets presented in \replace{Extended Data Figure 3a}{Supplementary Figure~\ref{sup:Extended data 3}a} to the master equation model, we accurately reproduce the hybridized charge qubit-resonator spectra, as shown in Fig.~\ref{fig:Fig4}b and \replace{Extended Data Figure 3b}{Supplementary Figure~\ref{sup:Extended data 3}b}.
\add{For this fit, the full 2D datasets are considered and a detuning dependence of the qubit decoherence rate $\Gamma(\varepsilon)$ is included in the model (see Methods for more details).}

From these spectra, we extract the \delete{renormalized }charge-photon coupling strengths \replace{$g_\mathrm{eff}$}{$g_0$}, and present them as a function of $f_\mathrm{r}$ in Fig.~\ref{fig:Fig4}e.
\replace{alongside the coupling strength $g_0 = g_\mathrm{eff} h f_\mathrm{r}/2t_\mathrm{c}$}
{We also estimate the effective charge-photon coupling strengths $g_\mathrm{eff}= g_0 2t_\text{c}/\sqrt {\varepsilon^2+4t_\text{c}^2} = g_0 2t_\mathrm{c}/h f_\mathrm{r}$, when the two systems are in resonance, and report them as a function of both $\varepsilon$ and $f_\mathrm{r}$ in Fig.~\ref{fig:Fig4}f}.
\replace{
Notably, while the evolution of $g_0$ does not exhibit a distinct trend within the measured frequency range, $g_\mathrm{eff}$ consistently decreases with increasing $f_\mathrm{r}$. 
This decrease corresponds to the anticipated quenching of the DQD electric dipole moment at finite detuning $\varepsilon$ due to charge localization (see Methods), resulting in $g_\mathrm{eff} = g_0 2t_\text{c}/\sqrt{\varepsilon^2+4t_\text{c}^2} \propto 1/f_\mathrm{r}$, as indicated by the black dashed line in Fig.~\ref{fig:Fig4}d \cite{2017_sto}. 
}{
Since $g_\mathrm{0} = \frac{1}{2}\beta_\mathrm{r}V_\mathrm{0,rms}/\hbar = \frac{1}{2}\beta_\mathrm{r}2\pi f_\mathrm{r}\sqrt{Z_\mathrm{r}/(2\hbar)}$, where the lumped equivalent resonator impedance $Z_\mathrm{r}$ can be written in terms of $f_\mathrm{r}$ as $Z_\mathrm{r} = 1/(2\pi f_\mathrm{r}C_\mathrm{r})$, we obtain a frequency dependence of $g_0 \propto \sqrt{f_\mathrm{r}}$ and hence $g_\mathrm{eff} \propto 1/
\sqrt{f_\mathrm{r}}$ (assuming constant $C_\mathrm{r}$ and resonance condition).
The dashed line in Fig.~\ref{fig:Fig4}e represents a fit of the extracted $g_0$ to the expected frequency dependence.
Using this fit, we estimate the evolution of $g_\mathrm{eff}$ as a function of $f_\mathrm{r}$ and illustrate it as a dashed line in Fig.~\ref{fig:Fig4}f.
Notably, the evolution of $g_0$ does not closely follow the expected trend as a function of $f_\mathrm{r}$. This discrepancy can be attributed to a nonuniform and frequency-dependent voltage profile of the resonator mode, potentially due to magnetic flux inhomogeneity along the SQUID array.
Alternatively, simultaneous hybridization of the DQD with higher order resonator modes (see Fig.~\ref{fig:Fig1}g), which approach the qubit frequency in the studied flux range, may influence the coupling to the fundamental mode. Further investigation is required in order to better understand the evolution of $g_0$.
}

\subsection*{Hybrid circuit QED with \replace{Wigner Molecules}{strongly correlated states}}

\begin{figure}[H]
     \centering
     \includegraphics[width=1\textwidth]{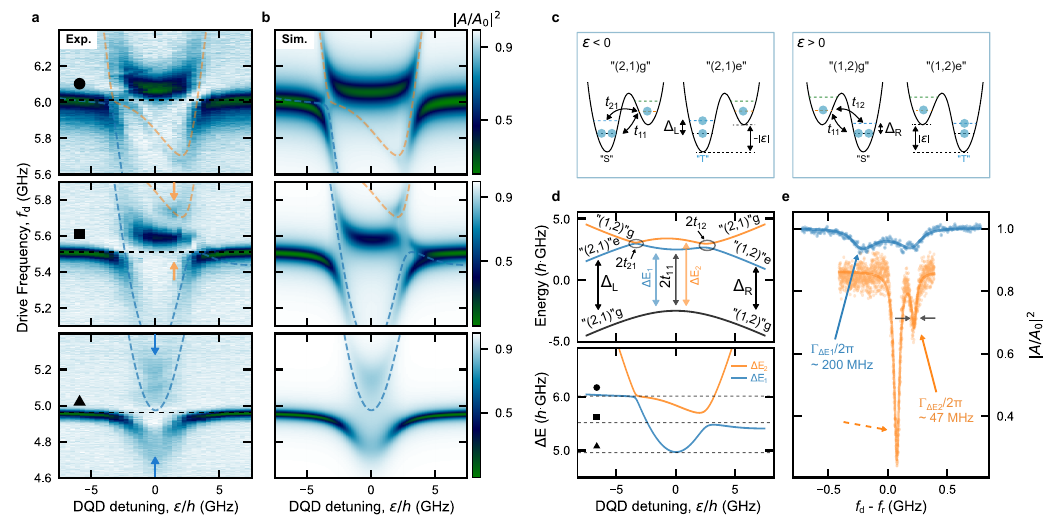}
     \caption{\textbf{Spectroscopy of the strongly correlated states in the hybrid architecture.} 
     \textbf{a,} Normalized amplitude of feedline transmission $|A/A_0|^2$ as a function of drive frequency $f_\mathrm{d}$ and DQD detuning $\varepsilon$. The three panels are taken in correspondence of three different bare resonator frequencies $f_\text{r}$ (black dashed lines denoted by black circle, square and triangle). 
     The dashed blue and orange lines show the calculated excitation spectra of the DQD, as detailed in panel \textbf{d}, revealing the presence of quenched \replace{WM states}{strongly correlated states (SCSs)}.
     \textbf{b,} Simulated $|A/A_0|^2$ using a generalized input-output theory of a multi-level DQD system (see Methods, and Supplementary Note~\ref{sup:io_wigner}) and for the three different $f_\text{r}$ as in \textbf{a}.
     The parameters for the simulations can be found in \replace{Extended Data Table 3}{Supplementary Table~\ref{sup:decoherence_matrix}}. 
     \textbf{c,} DQD schematics of the states relevant to ``$(2, 1)$" and ``$(1, 2)$" charge configurations for negative and positive DQD detuning $\varepsilon$. $\Delta_\mathrm{L}$ ($\Delta_\mathrm{R}$) is the singlet-triplet energy splitting $\Delta_\text{ST}$ when two holes are paired in the left (right) QD. 
     \textbf{d,} 
     Energy-level diagram (top panel) and excitation energy $\Delta E$ (bottom panel) calculated with the $4 \times 4$ Hamiltonian in Methods, and used for the input-output simulation in \textbf{b}. 
     In the bottom panel, the blue (orange) curve corresponds to the energy splitting $\Delta E_1$ ($\Delta E_2$) between the first (second) excited state branch and the ground state, shown in the top panel. 
     The black dashed lines denoted by a black circle, square, and triangle in the bottom panel represent the different $f_\text{r}$ used for acquiring the distinct spectra in \textbf{a}.
     \textbf{e,} Frequency line-cut taken at the DQD detuning indicated by the blue (orange) arrows in \textbf{a}. 
     The blue (orange) data highlights the resonator hybridization with the $\Delta E_1$ ($\Delta E_2$) transition.
     A fit to the master equation model (solid blue line), and Lorentzian (solid orange line) results in $\Gamma_{\Delta E_1}/2\pi \sim 200$~MHz and $\Gamma_{\Delta E_2}/2\pi \sim 47$~MHz, respectively.
     The orange dashed arrow indicates the resonator dispersively shifted by the interaction with the charge-like excitation $\Delta E_1$.
     }
     \label{fig:Fig5}
\end{figure}

Strikingly, our investigation of multiple adjacent inter-dot transitions reveals that the conventional charge qubit-like spectroscopy, as illustrated in Fig.~\ref{fig:Fig4}, featuring a single two-level system coupled to the resonator, fails to describe several cases. 
For instance, in Fig.~\ref{fig:Fig5}a, we present three independent measurements of the normalized feedline transmission \add{$|A/A_0|^2$} as a function of $f_\mathrm{d}$ and $\varepsilon$,
obtained for the same DQD configuration, but in correspondence to three different resonance frequencies $f_\mathrm{r}$ (indicated by black dashed lines denoted by a black circle, square, and triangle). 
See \replace{Extended Data Figure 4}{Supplementary Figure~\ref{sup:Extended data WM}} for a more detailed resonator spectroscopy.
These measurements unveil unconventional features, including anomalous spectroscopy diagrams asymmetric in $\varepsilon$, additional avoided crossings, and distinct spectroscopic lines
that deviate significantly from the conventional model for a resonator hybridized with a two-level system and have not been previously documented.

The anomalous spectrum of these DQD configurations is captured by an extended model that includes an excited state in each QD, whose \replace{energy is}{energies can be} close to $h f_\mathrm{r}$.
Specifically, we adopt a $4 \times 4$ Hamiltonian similar to the one used in prior studies \cite{2014_shi, 2012_shi} and numerically simulate the DQD spectrum and feedline transmission.
\add{We assume the validity of an effective hole numbering, where we neglect the even core holes (see Methods).} 
\add{More specifically, in modelling Fig.~\ref{fig:Fig5}a, we assume a $``(2,1)" \leftrightarrow ``(1,2)"$ DQD charge configuration (odd case).}
Within \replace{an effective}{the} $``(2,1)"$ configuration, the two holes in the left QD can occupy either the ground orbital state, forming $``(2,1)$g", or the \delete{exchange-split }excited orbital state, forming $``(2,1)$e" (\add{see }Fig.~\ref{fig:Fig5}c, \add{for }$\varepsilon < 0$).  
\add{Analogously, for $\varepsilon>0$, the eigenstates consist of the ground and excited states of the right QD, corresponding to $``(1,2)$g" and $``(1,2)$e", respectively (see Fig.~\ref{fig:Fig5}c, for $\varepsilon > 0$).}

\begin{figure}[H]
     \centering
     \includegraphics[width=1\textwidth]{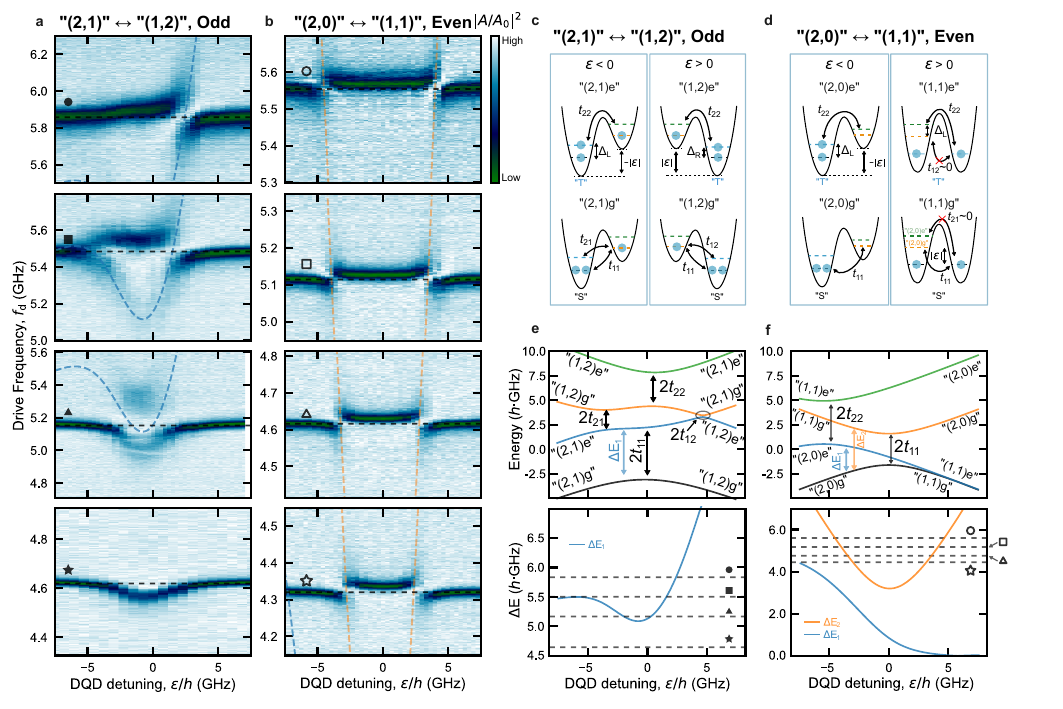}
     \caption{\textbf{Hole number parity-dependent behavior.} 
     \textbf{a, b,} DQD resonant spectra measured for two adjacent inter-dot transitions $``(2, 1)" \leftrightarrow ``(1, 2)"$ (odd, \textbf{a}) and  $``(2,0)$" $\leftrightarrow$ $``(1,1)"$ (even, \textbf{b}).
     Each panel presents the normalized amplitude of feedline transmission $|A/A_0|^2$ as a function of drive frequency $f_\mathrm{d}$ and DQD detuning $\varepsilon$, obtained in correspondence of the bare resonator frequency indicated by the horizontal black dashed line. 
     The dashed blue and orange lines show the calculated first, and second excitation spectra of the DQD, as detailed in panel \textbf{e} and \textbf{f}, revealing the presence of quenched \replace{WM states}{SCSs}.
     \textbf{c}, \textbf{d}, Schematics of the relevant states for $``(2, 1)" \leftrightarrow ``(1, 2)"$ (\textbf{c,} identical to the one shown in Fig.~\ref{fig:Fig5}c) and $``(2,0)$" $\leftrightarrow$ $``(1,1)"$ (\textbf{d}) inter-dot transition. 
     In the even parity case \textbf{d}, the energy gap between $``(1,1)$g" and $``(1,1)$e", $\Delta_\text{R} \sim 0$  due to the negligible exchange interaction of the unpaired holes. $\Delta_\text{L}$ is fixed to the value used in the odd case because the number of the holes in the left QD is unchanged.
     In the $``(1, 1)"$ charge state, ``S" (``T") denotes the spin singlet (triplet) state formed by the two holes in the respective QD. $t_\text{12}, t_\text{21} \sim 0$ due to the spin selection rule.
      \textbf{e, f,} 
     Energy level diagram (top panel) and excitation energy $\Delta E$ (bottom panel) calculated with the $4 \times 4$ Hamiltonian described in Methods, to obtain the resonant spectra in \textbf{a} and \textbf{b}, respectively.
     In the bottom panel, the blue (orange) curve corresponds to the energy splitting $\Delta E_1$ ($\Delta E_2$) between the first (second) excited state branch and the ground state, shown in the top panel. 
     The black dashed lines denoted by the different black filled (blank) symbols in the bottom panel of \textbf{e} (\textbf{f}) represent the different $f_\text{r}$ used for acquiring the distinct spectra in \textbf{a} (\textbf{b}).
     }
     \label{fig:Fig6}
\end{figure}

\add{
As we demonstrate in detail below, making use of Fig.~\ref{fig:Fig6}, such an effective particle numbering in the QDs readily captures spin structures that depend on the hole number parity \cite{2005_joh, 2021_jir, 2023_Grenoble}.
In this regard, we expect distinct spin symmetries in our $``(2,1)" \leftrightarrow ``(1,2)"$ configuration related to the ground and excited states \cite{2012_shi}.
More explicitly, we assume the ground (excited) state to involve anti-symmetric singlet~``S" (symmetric triplet~``T") spin pairing in the doubly-occupied QD (Fig.~\ref{fig:Fig5}c).
In this configuration, the two lowest energy levels form doublet spin states together with the single spin in the other QD \cite{2012_shi}.
For instance, $``(2,1)$g" ($``(2,1)$e") forms a doublet state with spin singlet (triplet) pairing in the left QD. 
Here, finite exchange interaction can couple the ground and excited doublet states \cite{2012_shi,2014_shi},
because they have the same total spin quantum number $S_\mathrm{tot}=1/2$, despite the different spin symmetries within the doubly-occupied QD.
With this exchange interaction, the $``(2,1)\mathrm{g}" \leftrightarrow ``(2,1)$e" or $``(1,2)\mathrm{g}" \leftrightarrow ``(1,2)$e" transitions can be revealed by our resonator as presented in Fig.~\ref{fig:Fig5}a, in agreement with the spin selection rule.
}

Based on the above modeling, we empirically determine the Hamiltonian parameters,
including $\Delta_\mathrm{L}/h = 5.40$~GHz ($\Delta_\mathrm{R}/h = 4.73$~GHz), i.e. the \add{singlet-triplet splitting} $\Delta_\mathrm{ST}$ in the left (right) QD, 
that accurately reproduce both the energy and excitation spectra, reported respectively in the top and bottom panels of Fig.~\ref{fig:Fig5}d \add{(see Supplementary Table~\ref{sup:fig5_parameters})}.
We also estimate the tunnel coupling strengths between the $i^\text{th}$ state of the left QD and the $j^\text{th}$ state of the right QD, $t_\mathrm{ij}$.
Furthermore, we use input-output theory to analyze the interaction between the resonator and the multi-level QD system \cite{2016_bur} \add{(see Methods and Supplementary Note~\ref{sup:io_wigner})},
enabling us to accurately reproduce the spectrum of the hybridized system, as depicted in Fig.~\ref{fig:Fig5}b \delete{(see Methods and Supplementary Note~\ref{sup:io_wigner})}.
\add{Note that for this, it is essential to assume finite tunnel coupling strengths $t_{12}/h = 0.21$~GHz and $t_{21}/h = 0.11$~GHz.}

The extracted values of $\Delta_\mathrm{L,R}$ are orders of magnitude smaller than the expected orbital energy gap ($\sim$~70$~h\cdot$GHz) \replace{based on}{obtained from a single particle model, considering} the dimensions of our QDs \add{(see Supplementary Note~\ref{sup:wigner})}.
Instead, these \replace{extracted}{estimated} excitation energies \replace{are consistent with Wigner molecularization}{are generated by strong Coulomb correlation effects} within each QD.
To support this interpretation, in Supplementary Note~\ref{sup:wigner}, we present a preliminary model for two interacting holes in planar Ge, which suggests that the small anisotropy in \replace{our}{the} QD confinement, in conjunction with electron-electron interactions, can result in \replace{WM}{strongly correlated states (SCSs)} with $\Delta_\mathrm{ST}\lesssim 10~h\cdot$GHz.
Although a more comprehensive investigation based on full-configuration-interaction calculations is necessary to precisely characterize the energy scales within the DQD \cite{2021_cor, 2022_yan}, our preliminary analysis provides \delete{clear} evidences that the observed features in Fig.~\ref{fig:Fig5}a \replace{are related}{may be attributed} to WM \replace{s}{ states} \cite{2022_yana, Erc_2021, 2022_yan, 2007_yan, 1999yannouleas, 2009li, 2004szafran}.

\delete{
To bolster our assertion,}
\add{As described above,} we note that \replace{WM states}{SCSs} exhibit distinct spin symmetries, with the ground orbital state supporting the anti-symmetric spin singlet, and the excited orbital state supporting the symmetric spin triplet \cite{2012_shi,2023_jan} (see Fig.~\ref{fig:Fig5}c).
These symmetries imply that the relaxation \replace{process from the upper energy branch to the ground branch}{between the states specified above} involves a spin-changing process, 
which can be considerably slower compared to the bare charge relaxation \delete{processes} \cite{2012_shi, 2023_jan}. 
We explore this distinction in Fig.~\ref{fig:Fig5}e, which presents two line-cuts along $\varepsilon$, marked by the orange and blue dashed lines in the middle and bottom panels of Fig.~\ref{fig:Fig5}a.
In comparison to the charge qubit-like decoherence rate $\Gamma_{\Delta E_1}/2\pi \sim 200$~MHz extracted from the fit to the master equation (blue solid line in Fig.~\ref{fig:Fig5}e), the second excited state spectrum \add{$\Delta E_2$} (denoted by an orange solid arrow in Fig.~\ref{fig:Fig5}e) is characterized by a significantly narrower linewidth $\Gamma_{\Delta E_2}/2\pi \sim 47$~MHz (extracted from the fit to the Lorentzian), \replace{providing further evidence supporting}{which is further supporting our modeling.} 
\replace{This asymmetry in DQD-resonator}{Similar} spectroscopic signatures, attributed to the \replace{WM states}{SCSs} with excitation energies very close to that of the resonator, have been detected across multiple inter-dot transitions.
Supplementary Figure~\ref{sup:Extended data another WM} reports another instance of a similar spectrum
exhibiting $\Gamma_{\Delta E_2}/2\pi \sim 35$~MHz.

\add{The presented Hamiltonian 
also models a spin-charge hybrid qubit, which can be encoded in the SCSs exhibiting a lower decoherence rate in comparison to a bare charge qubit.
Such a hybrid qubit also allows all-electrical control of the spin states based on exchange interaction \cite{2012_shi, 2021_jan}.
}

To further explore these unconventional DQD spectra \replace{in relation to the creation of WM,}
{due to strong Coulomb correlation effects, and to confirm the intrinsic spin nature of the aforementioned states,}
we delve into the expected hole number parity-dependence, distinguishing between even and odd effective DQD occupation.
\add{As we detail below, the observed energy spectra measured with our resonator both in the even and odd configurations are consistent with the spectra derived from our effective model, which takes into account the parity dependent spin structures.}
In Figs.~\ref{fig:Fig6}a and b, we investigate \add{a representative instance of} two neighboring inter-dot transitions involving effective charge configurations $``(2, 1)" \leftrightarrow ``(1, 2)"$ and $``(2, 0)" \leftrightarrow ``(1, 1)"$, respectively (\replace{see}{denoted in the stability diagram in} Supplementary Figure~\ref{sup:Fig_EvenOdd_Stab}a \replace{for the corresponding stability diagram}{by the dashed black and red boxes}).

In the \replace{$``(2, 1)"$}{$``(2, 1)" \leftrightarrow ``(1, 2)"$} configuration (Fig.~\ref{fig:Fig6}a), characterized by an odd total number of holes, the ground and first excited states have the same spin quantum number \add{$S_\mathrm{tot}=1/2$}. 
\delete{Consequently, finite tunneling rates $t_{21}$ and $t_{12}$ exist between them (see Figs.~\ref{fig:Fig6}c, e)}
\add{Similar to the configuration shown in Fig.~\ref{fig:Fig5}},
this results in a finite exchange interaction between $``(2,1)$g" and $``(2,1)$e", \add{and between $``(1,2)$g" and $``(1,2)$e",} enabling their electrical coupling to the resonator, in accordance with the spin selection rule.
To faithfully replicate both energy and excitation spectra shown in Fig.~\ref{fig:Fig6}e,
it is essential to assume a sizable
\replace{tunnel coupling between different exchange-split states within the respective QDs ($t_{21}$)}{$t_{21}$} and a relatively small $\Delta_\mathrm{L}/h \sim 5.48$~GHz.
\add{In contrast to Fig.~\ref{fig:Fig5}a, the resonator spectroscopy reported in Fig.~\ref{fig:Fig6}a does not show the second excitation $\Delta E_2$, due to the larger $\Delta_\mathrm{R}$ with respect to $f_\mathrm{r}$.
}

For the adjacent even configuration, denoted as $``(2,0)$" $\leftrightarrow$ $``(1,1)"$ (see Figs.~\ref{fig:Fig6}b and d), with a \add{single} additional hole in the right QD compared to the odd configuration, the total spin numbers of the ground (\replace{singlet}{$S_\mathrm{tot}=0$}) and first excited states (\replace{triplet}{$S_\mathrm{tot}=1$}) in the DQD are different.
\add{This is further supported by the observed signatures of Pauli spin blockade (PSB) presented in Supplementary Note~\ref{sup:even_odd}.
Because the number of holes in the left QD is the same as in the configuration presented in Fig.~\ref{fig:Fig6}a, we expect a similar $\Delta_\mathrm{L}$ between the ``(2,0)g" and ``(2,0)e" states.
The extracted width of the PSB window $w_\mathrm{PSB}/h \sim 7.8 \pm 1.2$~GHz is comparable to $\Delta_\mathrm{L}/h \sim 5.48$~GHz used for the simulation of the energy diagram in the odd configuration in Fig.~\ref{fig:Fig6}e.
}
\replace{
Consequently, a transition between these states requires magnetic interactions, such as spin-orbit coupling or g-factor anisotropies, both of which are negligible in our experimental setup ($B\ll$ 1 mT) \mbox{\cite{2015_nic, 2021_jir}}.
As a result, in modeling the even configuration, we set the tunneling rates $t_{12}, t_{21} \sim 0$, resulting in a negligible coupling of the ground-excited state transition to the resonator.}
{In our system with a magnetic field $B\ll1$~mT, the effect of the spin-orbit interaction or Zeeman splitting difference between the two QDs can be neglected \cite{2021_jir, 2022_gey}. 
Consequently, $t_{12}, t_{21} \sim 0$ in the model Hamiltonian, and the resonator electric field can only drive spin-preserving transitions.}
Furthermore, in the $``(1,1)"$ configuration, the spatial separation of the two holes results in a negligible exchange splitting between the ``(1,1)g" and ``(1,1)e" states (Fig.~\ref{fig:Fig6}d), allowing us to set $\Delta_\mathrm{R} \sim 0$ in the model Hamiltonian.
This leads to the energy \replace{level spectrum}{diagram} depicted in Fig.~\ref{fig:Fig6}f\replace{As a result, Fig.~\ref{fig:Fig6}b exhibits a}
{, which explains the observation of a}
conventional charge-qubit-like spectrum \add{reported in Fig.~\ref{fig:Fig6}b}, corresponding to the charge transition \delete{between} $``(2, 0)\mathrm{g}" \leftrightarrow ``(1, 1)\mathrm{g}"$ \add{with excitation energy $\Delta E_2$}.
We also demonstrate that the master equation model constructed using our extended effective Hamiltonians closely reproduces the measured spectra of the hybridized multi-level DQD-resonator system (Figs.~\ref{fig:Fig6}a and b) for both odd and even configurations, as reported in Supplementary Note~\ref{sup:even_odd_spectroscopy}.


\add{As a side note, we observe faint additional features in some of our 2D spectroscopy datasets, such as the ones around 5~GHz in Fig.~\ref{fig:Fig6}a. These features may be attributed to uncontrolled two-level fluctuators in the tunneling junctions of the SQUID array resonator, which can capacitively couple to the microwave photons (see Supplementary Note~\ref{sup:cav_char}) \cite{2019_Mul}. Alternatively, transitions between higher DQD energy levels, which are observable due to a finite thermal population of the excited state, might also explain some of these extra avoided crossings. However, accurately modeling these transitions would require the introduction of additional energy states into our model, which is beyond the scope of this work.}

\section{Discussion}

In this study, we have demonstrated the potential of a hybrid architecture, combining a superconductor cavity with semiconductor QDs for advancing hole-based quantum information processing in planar germanium.
Leveraging a high-impedance Josephson junction-based resonator with tunable frequency, we have demonstrated strong hole charge-photon coupling. 
This achievement is substantiated by our observation of charge-photon vacuum-Rabi mode splitting and the high cooperativity value ($C \sim 100$) estimated for our hybrid system. 
Furthermore, the frequency tunability of our resonator has enabled us to successfully resolve \replace{strongly correlated Wigner molecule states}{strongly correlated states (SCSs)} within QDs in planar Ge structures.
The distinct spin \replace{structure}{symmetries} of the \replace{Wigner molecule spectrum}{SCSs} lead to significantly reduced decoherence rates of the higher excited levels, a promising development for establishing strong spin-photon coupling. 
The interaction between QD \replace{Wigner molecules}{SCSs} and a frequency-tunable resonator provides a very effective avenue for exploring complex many-body electronic states in multi-level QDs.
\add{While a detailed measurement of the charge density distribution of the ground state is required to unambiguously prove the Wigner molecularization process \cite{2019_sha}, the presence of strong Coulomb correlation and QD confinement anisotropy, as suggested by our simulations, make Wigner molecules the most plausible model to describe the observed quenching of strongly correlated states in our QDs \cite{2022_yana, Erc_2021, 2022_yan, 2007_yan, 1999yannouleas, 2009li, 2004szafran}}.
Our findings facilitate coherent photon coupling with spin-charge hybrid qubits, 
\replace{but also support in-depth studies of strongly correlated electronic states in open systems.}
{also potentially based on longitudinal interaction through singlet-triplet splitting modulation \cite{2021abadillo-uriel}.}
In conclusion, we have demonstrated the ability to coherently exchange a photon with holes in planar Ge, marking a critical step toward achieving long-distance spin-spin entanglement. 
Our work lays the foundation for future research on hole-photon coupling and long-range interactions of hole-based qubits, paving the way for the development of large-scale quantum processors.

\newpage

\section{Methods}
\subsection*{Device fabrication}
The hybrid triple QD device is fabricated on a Ge/SiGe heterostructure grown by low-energy plasma-enhanced chemical vapor deposition (LEPECVD) using a forward grading technique (see Fig.~\ref{fig:Fig1}b) \cite{2021_jir}. The device fabrication is entirely carried out at the Center of MicroNano Technology (CMi) at EPFL. As a first step, 60 nm Pt markers and ohmic contacts are patterned by E-beam lithography (EBL), evaporation and lift-off. Immediately before the deposition, a 20~s dip in diluted HF (1\%) removes the native oxide in the opened regions to ensure a low-resistive ohmic contact. The two-dimensional hole gas (2DHG) is self-accumulated in the 16 nm Ge quantum well (QW). Therefore, a 110 s reactive ion etching (RIE) step etches $\approx$ 80-90 nm, leaving a well defined conductive channel from one ohmic contact to the other. The reacting plasma is based on SF$_6$, CHF$_3$ and O$_2$ and the mask is patterned by EBL. A 15~s dip in buffered HF (BHF) etches away the native oxide immediately before the gate oxide deposition, a 20~nm atomic layer deposition (ALD) Al$_2$O$_3$. The deposition temperature is 300$^\circ$C. Then, a 15 min rapid thermal annealing (RTA) in forming gas (N$_2$/H$_2$ 5\%) at 300$^\circ$C \replace{decreases the channel sheet resistance to $\approx 100\ \Omega$/sq.}{ensures that the Pt properly diffuses down to the Ge QW}. The single-layer gates are patterned in two steps by EBL, evaporation and lift-off. This ensures that the thin 3/22~nm Ti/Pd gates are patched on the etched step by 3/97~nm ones, routed out to the bonding pads. The superconducting part of the device is again patterned in two steps by EBL, evaporation and lift-off. First, the waveguide and the ground plane (120~nm of Al) and, lastly, the SQUID array resonator, following the conventional Dolan-bridge double angle evaporation method for Josephson junctions (JJs). The bottom Al layer is 35~nm thick, whereas the top one is 130~nm. The tunneling oxide barrier is grown by filling the chamber with O$_2$ at a pressure of 2~Torr for 20~min (static oxidation) without breaking the vacuum.     
From measurements of the SQUID array resistance at room temperature, we estimate a critical current of about 522~nA per SQUID.

\subsection*{Fitting procedure for a conventional cavity-dressed charge qubit}

The experimental data shown in this work reporting the feedline transmission $S_\mathrm{21}$ are fitted to a master equation model (see Supplementary Note~\ref{sup:io_cq} for the full derivation) and normalized by a background trace to remove the standing wave pattern present in the feedline transmission. The background reference trace is obtained by tuning the resonance frequency of the resonator outside the frequency region of interest by making use of the superconducting coil and recording a high-power trace.
The complex transmission of a resonator hanged to a 50 $\Omega$ feedline and coupled to a charge qubit reads as:
\begin{equation}
     S_{21} = ae^{i\alpha}e^{-2\pi if_\mathrm{d}\tau} \ \frac{\Delta_r-i(\kappa-|\kappa_\mathrm{ext}|e^{i\phi})/2+g_\mathrm{eff}\chi}{\Delta_r-i\kappa/2+g_\mathrm{eff}\chi},
     \label{eq:master_equation}
\end{equation}
where $\Delta_r=\omega_\mathrm{r}-\omega_\mathrm{d}$ is the resonator-drive detuning, $\kappa=\kappa_\mathrm{ext}+\kappa_\mathrm{int}$ the total resonator linewidth given by both coupling to the waveguide $\kappa_\mathrm{ext}=|\kappa_\mathrm{ext}|e^{i\phi}$ and internal losses $\kappa_\mathrm{int}$, $g_\mathrm{eff}=g_0\frac{2t_c}{\sqrt{\varepsilon^2+4t_c^2}}$ the effective charge-photon coupling strength, $\varepsilon$ the DQD detuning, $t_\mathrm{c}$ the inter-dot tunneling coupling, $\chi=\frac{g_\mathrm{eff}}{-\Delta_\mathrm{q}+i\Gamma}$ the DQD susceptibility, $\Delta_\mathrm{q}= \omega_\mathrm{q}-\omega_\mathrm{d}$ the qubit-drive detuning\add{, with the qubit frequency $\omega_\mathrm{q}/2\pi=\sqrt{\varepsilon^2+4t_\mathrm{c}^2}/h$,} and $\Gamma$ the charge qubit linewidth. $a$, $\alpha$, $\tau$, and $\phi$ are correction factors that take into account the non-ideal response of the cavity due to the environment. Further information is provided in Supplementary Note~\ref{sup:io_cq}.
The resonator parameters $f_\mathrm{r}$, $\kappa$ and $\kappa_\mathrm{ext}$, as well as the environmental factors, are obtained by separately fitting $S_\mathrm{21}$ for the bare uncoupled resonator.

The simultaneous fit of the line-cuts reported in \replace{Figs. \ref{fig:Fig2}g - i}{Fig.~\ref{fig:Fig2}d and Supplementary Figures~\ref{sup:Fig2}g - i} is performed using common fitting parameters for $t_\mathrm{c}$ and $\Gamma$, while using separate $g_0^k$, where $k=$ g, h, i for the three different datasets.
In order to convert the voltage axis to DQD detuning $\varepsilon = \mu_\mathrm{L} - \mu_\mathrm{R} = \beta_\mathrm{pL}V_\mathrm{pL} - \beta_\mathrm{pR}V_\mathrm{pR}$, the differential lever arms $\beta_\mathrm{pL}=\alpha_\mathrm{pL}^L - \alpha_\mathrm{pL}^\mathrm{R} = 0.031$
and $\beta_\mathrm{pR}=\alpha_\mathrm{pR}^\mathrm{R} - \alpha_\mathrm{pR}^\mathrm{L} = 0.016$ are extracted from Coulomb diamond and DQD charge stability diagram measurements. Here, $\mu_\mathrm{L}$ ($\mu_\mathrm{R}$) is the electrochemical potential of the left (right) QD, $\alpha_\mathrm{pL}^\mathrm{L}$ ($\alpha_\mathrm{pR}^\mathrm{R}$) is the lever arm for the left (right) plunger gate
and $\alpha_\mathrm{pL}^\mathrm{R}$ ($\alpha_\mathrm{pR}^\mathrm{L}$) is the cross-lever arm for the left (right) plunger gate.

\delete{
While Fig.~\ref{fig:Fig3}b is a frequency line-cut of Fig.~\ref{fig:Fig3}a at the resonance point,}
Fig.~\ref{fig:Fig3}c is obtained from a separate measurement \add{with respect to Fig.~\ref{fig:Fig3}a}, with higher resolution and integration time \add{and taken at a slightly different flux point ($V_\mathrm{flux}=507$~mV)}.
The resonator parameters used for generating Fig.~\ref{fig:Fig3}d are obtained from fitting the bare resonator as a function of \replace{the magnetic flux}{$V_\mathrm{flux}$}, similar to the measurement in Fig.~\ref{fig:Fig1}g
\add{(see Supplementary Figure~\ref{sup:Suppl_kappas_vs_fr}).
The other parameters are obtained from fitting a frequency line-cut of Fig.~\ref{fig:Fig3}b at $V_\mathrm{flux}=504$~mV (see Supplementary Figure~\ref{sup:fig3_flux_sweep}b), where the two subsystems are in resonance, to the master equation model described above.
}
Note that here, the lowest frequency of the measurement is 3.8~GHz, limited by the bandwidth of the cryogenic circulators.
To help interpret the different extracted $g_0$ for the datasets in Fig.~\ref{fig:Fig3}a and Supplementary Figure~\ref{sup:ext_fig2}, the resonator differential lever arm is calculated for both cases, following the relation $g_0=\frac{1}{2}\beta_\mathrm{r} V_\mathrm{0,rms}/\hbar = \frac{1}{2}\beta_\mathrm{r}2\pi f_\mathrm{r}\sqrt{\frac{Z_\mathrm{r}}{2\hbar}}$ 
\cite{2023_Grenoble}. 
Using $Z_\mathrm{r}=1.6$  $\mathrm{k}\Omega$ (1.2 k$\Omega$) and $f_\mathrm{r}=4.149$~GHz (5.432~GHz) for Fig.~\ref{fig:Fig3}a (\replace{Fig.~3d}{Supplementary Figure~\ref{sup:ext_fig2}a}), we get $V_\mathrm{0,rms} = 7.6~\mathrm{\mu}$V ($8.7~\mathrm{\mu}$V) and $\beta_\mathrm{r}=0.18$ eV/V (0.25 eV/V). For further details about the resonator equivalent lumped impedance and its frequency, see Supplementary Notes~\ref{sup:cav_char}.

\replace{
The fits presented in Fig.~\ref{fig:Fig4}b and Extended Data Figure 3 are performed on the entire 2D datasets (see panels a of relative figures) rather than to single line-cuts.
$g_\mathrm{eff}$ is used as a fitting parameter, while $g_0$ is estimated using $g_0=g_\mathrm{eff}f_\mathrm{q}/2t_\mathrm{c}$.
Here, $f_\mathrm{q}=f_\mathrm{r}$ when the two systems are in resonance.
The inter-dot tunnel coupling rate $t_\mathrm{c}$ is only fitted for the lowest frequency measurement and kept constant for subsequent fits.
Additionally, the differential lever arm $\beta_\mathrm{pL}$ of the left plunger gate is used as a fitting parameter and the obtained fitted values are in good agreement with the values extracted from DQD stability diagram measurements.}
{
In contrast to Fig.~\ref{fig:Fig3}, the fits presented in Fig.~\ref{fig:Fig4} and Supplementary Figure~\ref{sup:Extended data 3} are performed simultaneously on all the 2D spectroscopy datasets.
To account for the detuning dependence of the qubit dephasing rate due to charge noise,
we include a DQD detuning dependence of the qubit decoherence in the form of $\Gamma = \Gamma_0 + \Gamma_\varepsilon \frac{1}{\hbar} \frac{\partial\omega_\mathrm{q}}{\partial\varepsilon} = \Gamma_0 + \Gamma_\varepsilon \frac{\varepsilon}{\hbar\omega_\mathrm{q}}$ \cite{2019_sca,2014_pal,2017_tho},
where the derivative $\frac{\partial\omega_\mathrm{q}}{\partial\varepsilon}$ quantifies the sensitivity of the qubit energy, and hence the scaling of the qubit dephasing rate,
with respect to detuning noise induced by charge noise in the environment.
For the combined fit, the DQD tunnel coupling $t_c$, the differential lever arm $\beta_\mathrm{pL}$ of the left plunger gate as well as the constant and detuning-dependent decoherence coefficients $\Gamma_0$ and $\Gamma_\varepsilon$, respectively, are shared among all five datasets, while the charge-photon coupling strength $g_0$, a voltage offset $V_\mathrm{pL}^0$ that corresponds to $\varepsilon=0$ and all the resonator parameters are fitted independently for each dataset.
$\Gamma_0$ was fixed to 57~MHz, extracted from the fit in Fig.~\ref{fig:Fig3}c. However, we verified that this does not have any influence on the estimated values for $g_0$. \adds{We extract a value of $\Gamma_\mathrm{\varepsilon} \sim$ 164 MHz from the fit.}
All resonator parameters (see Eq.~\eqref{eq:master_equation}), except for the bare resonator frequency $f_\mathrm{r}$, are estimated by fitting a single trace of $S_{21}$ taken at large DQD detuning $\varepsilon$. 
The error bars in Figs.~\ref{fig:Fig4}e and f correspond to the $2\sigma$ confidence interval estimated by propagating the errors of $\kappa$ and $\kappa_\mathrm{ext}$ taken from the separate resonator fit.
The dashed line in Fig.~\ref{fig:Fig4}e represents a fit (including the errors of $g_0$) of the extracted $g_0$ values to the relation $g_0 = a \cdot \sqrt{f_\mathrm{r}}$.
The resulting evolution of $g_0$ is then converted to $g_\mathrm{eff}=g_02t_\mathrm{c}/f_\mathrm{r}$ and reported in Fig.~\ref{fig:Fig4}f.
The shaded regions in Figs.~\ref{fig:Fig4}e and f correspond to the $2\sigma$ confidence interval extracted from the last fit above.
}

\subsection*{Hamiltonian for \replace{WM}{SCS} simulation}

To numerically reproduce the hybridized DQD-resonator spectra obtained from the microwave feedline transmission 
shown in Figs.~\ref{fig:Fig5},~\ref{fig:Fig6}, Supplementary Note~\ref{sup:WM_details} and Supplementary Note~\ref{sup:even_odd_spectroscopy}, we need, as a first step, to identify the multi-level energy spectra characterizing the DQD in each configuration.

We model the DQD assuming a 4 $\times$ 4 toy-model Hamiltonian identical to the spin-charge hybrid qubit defined by three-particles in a DQD \cite{2012_shi, 2014_shi}, as reported below. The Hamiltonian is written in the position basis $[|\mathrm{L}_\text{g}\rangle, |\mathrm{L}_\text{e}\rangle, |\mathrm{R}_\text{g}\rangle, |\mathrm{R}_\text{e}\rangle]$, where L (R) denotes the charge state with the excess hole in the left (right) QD. g and e present the ground and excited states of the corresponding charge configuration, respectively.
Specifically, in the case of an odd total number of holes in the DQD, L = (2n + 2, 2m + 1) and R = (2n + 1, 2m + 2). 
Here, we use the notation (p, q) to denote the DQD charge number configuration with p (q) representing the number of holes in the left (right) QD.
Throughout this work, the 2n (2m) core holes in the left (right) QD play no role, reducing the effective DQD charge number to L = ``(2, 1)" and R = ``(1, 2)", respectively. Similarly, in the even configuration, the relevant charge states effectively become L = ``(2, 0)", and R = ``(1, 1)".  
Consequently, the basis in which the Hamiltonian is expressed is [$``(2, 1)\mathrm{g}"$, $``(2, 1)\mathrm{e}"$, $``(1, 2)\mathrm{g}"$, $``(1, 2)\mathrm{e}"$] in the odd configuration and [$``(2, 0)\mathrm{g}"$, $``(2, 0)\mathrm{e}"$, $``(1, 1)\mathrm{g}"$, $``(1, 1)\mathrm{e}"$] in the even one (see Figs.~\ref{fig:Fig6}c and d for the schematic visualisation of these states).
In this basis, the DQD $4 \times 4$ Hamiltonian reads:

\begin{equation}
H = 
\begin{bmatrix}
\varepsilon/2 & 0 & t_\text{11} & t_\text{12} \\
0 & \eta_\text{L}\varepsilon/2 + \Delta_L & t_\text{21} & t_\text{22} \\
t_\text{11} & t_\text{21} & -\varepsilon/2 & 0 \\
t_\text{12} & t_\text{22} & 0 & -\eta_\text{R}\varepsilon/2 + \Delta_R \\
\end{bmatrix}\label{eq:HWM}
\end{equation}

Here, $\varepsilon$ is the DQD detuning, $\Delta_\text{L} (\Delta_\text{R})$ is the singlet-triplet splitting $\Delta_\mathrm{ST}$  \add{when two holes are paired} in the left (right) QD and $t_\text{ij}$ denotes the tunnel coupling between the $i^\text{th}$ state of left QD and $j^\text{th}$ state of right QD. We also include $\eta_\text{L} = 0.92 \; (\eta_\text{R} = 0.913)$ to account for the different lever arms of the excited  states in Fig.~\ref{fig:Fig5} \cite{2022_yan}. 

The Hamiltonian eigenvalues are used to reconstruct the energy spectra (eigenenergies vs $\varepsilon/h$) reported in Figs.~\ref{fig:Fig5}d and \ref{fig:Fig6}e, f (top panels), whereas the excitation spectra, i.e. the energy differences between excited states and the ground state, are displayed in the bottom respective panels.

In the odd total hole number configuration, $``(2, 1)\mathrm{k}" \leftrightarrow ``(1, 2)\mathrm{k}"$, with k = g, e (Figs.~\ref{fig:Fig5} and Figs.~\ref{fig:Fig6}a), the ground and the first excited states have the same total spin number $S_\mathrm{tot}=1/2$. For example $``(2, 1)\mathrm{g}"$ and $``(2, 1)\mathrm{e}"$ form the doublet spin states with an energy splitting given by the exchange interaction of the paired holes in the left QD \cite{2012_shi}.
Thereby, a finite tunnel coupling between ground and exchange-split excited states is allowed, e.g. $t_{12}, t_{21} > 0$, by spin-selectrion rules \cite{2012_shi}. Both the QDs can have $\Delta_\mathrm{ST}/h\sim 5$~GHz (close to resonator frequency) when a \replace{WM}{SCS} is formed in each QD.

In contrast, in the even configuration, $``(2, 0)\mathrm{k}" \leftrightarrow ``(1, 1)\mathrm{k}"$ (Fig.~\ref{fig:Fig6}b), the ground ($S_\mathrm{tot}=0$) and first excited ($S_\mathrm{tot}=1$) states do not have the same spin quantum number \cite{2005_joh}. 
For this reason, 
the terms $t_{12}, t_{21} \sim 0$, the transition rates for $``(2, 0)\mathrm{e}" \leftrightarrow ``(1, 1)\mathrm{g}"$ (corresponding to $``(2, 0)\mathrm{T}" \leftrightarrow ``(1, 1)\mathrm{S}"$) and $``(2, 0)\mathrm{g}" \leftrightarrow ``(1, 1)\mathrm{e}"$ (corresponding to $``(2, 0)\mathrm{S}" \leftrightarrow ``(1, 1)\mathrm{T}"$) are negligible in our setup with $B \ll 1$ mT (see Fig.~\ref{fig:Fig6}d) \cite{2015_nic, 2021_jir}. 
Also, \replace{due to the negligible exchange interaction between the unpaired holes in the $``(1, 1)\mathrm{k}"$ configuration,}
{
spatial separation of the holes in the $``(1, 1)"$ configuration, results in a negligible exchange splitting between $``(1, 1)\mathrm{g}"$
and $``(1, 1)\mathrm{e}"$, and}
we set $\Delta_\text{R} = 0$ in model Hamiltonian for the even case. 
Because the number of holes in the left QD is the same as in the odd case, we keep the same value of $\Delta_\text{L}$ as in the odd case.

The aforementioned Hamiltonian, combined with the generalized input-output theory for a multi-level DQD system interacting with a superconducting resonator (see  Supplementary Note~\ref{sup:io_wigner}) \cite{2016_bur}, reproduces the features observed in the panels of Fig.~\ref{fig:Fig5}a and Figs.~\ref{fig:Fig6}a and b. The relevant Hamiltonian parameters are shown in Supplementary Note~\ref{sup:WM_details} and Supplementary Note~\ref{sup:even_odd_spectroscopy}.

It is worth noting that a similar model can be applied to investigate a resonator coupled to a generic multi-level DQD system, thus offering opportunities to explore valley-orbit states in silicon coupled to superconducting resonators \cite{2021_bor}.

\newpage

\section{Data availability}
The datasets generated during the current study are available in Zenodo with the identifier \url{https://doi.org/10.5281/zenodo.13935167}.

\backmatter

\newpage


\newpage
\section*{Acknowledgements}
The authors thank Simone Frasca, Vincent Jouanny, Guillaume Beaulieu, Camille Roy, Dominic Dahinden, Davide Lombardo, Daniel Chrastina and Siddhart Gautam for contributing in some cleanroom fabrication steps, the measurement setup, device simulations, data analysis and for the useful discussions.

P.S. acknowledges support from the Swiss National Science Foundation (SNSF) through the grants Ref. No. 200021 200418 and Ref. No. 206021\_205335, and from the Swiss State Secretariat for Education, Research and Innovation (SERI) under contract number 01042765 SEFRI MB22.00081.
W.J. acknowledges support from EPFL QSE Postdoctoral Fellowship Grant.
S.B, D.L and P.S acknowledge support from the NCCR Spin Qubit in Silicon (NCCR-SPIN) Grant No. 51NF40-180604.
M.J., G.K., G.I. and S.C. acknowledge support from the  Horizon Europe Project IGNITE ID 101070193.
G.K. acknowledges support from the FWF via the P32235 and I05060 projects.

\section*{Contributions}
F.D.P., F.O., W.J. and P.S. conceived the project.
F.D.P. and F.O. fabricated the device and built the experimental setup. 
F.D.P., F.O. developed the fabrication recipe with inputs from M.J. and G.K..
F.D.P., F.O. and M.J. designed the hybrid device.
F.D.P., F.O. and W.J. performed the electrical measurements and analyzed the data. 
W.J. and S.B. derived the theoretical model and simulated the \replace{Wigner molecule}{strongly correlated} states. 
G.I. and S.C. designed the SiGe heterostructure and performed the growth. 
P.S., D.L. and G.K. initiated the project.
P.S., D.L. supervised the project.
F.D.P., F.O., W.J. and P.S. wrote the manuscript with inputs from the authors.
F.D.P., F.O., W.J. contributed equally to this work.

\section*{Competing interests}
The authors declare no competing interests.

\clearpage
\setcounter{section}{0}
\setcounter{figure}{0}
\setcounter{table}{0}
\setcounter{tocdepth}{-1}


\renewcommand{\figurename}{Supplementary Figure}
\renewcommand{\tablename}{Supplementary Table}
\renewcommand\refname{Supplementary References}

\begin{center}
    \textbf{\Large Supplementary Information: Strong hole-photon coupling in planar Ge for probing charge degree and strongly-correlated states}
\end{center}
\vspace{1em}

\addtocontents{toc}{\setcounter{tocdepth}{2}}

{\hypersetup{linkcolor=black}
	\tableofcontents
}

\section{Experimental Setup}\label{supplsec1}
\begin{figure}[H]
     \centering
     \includegraphics[width=150mm]{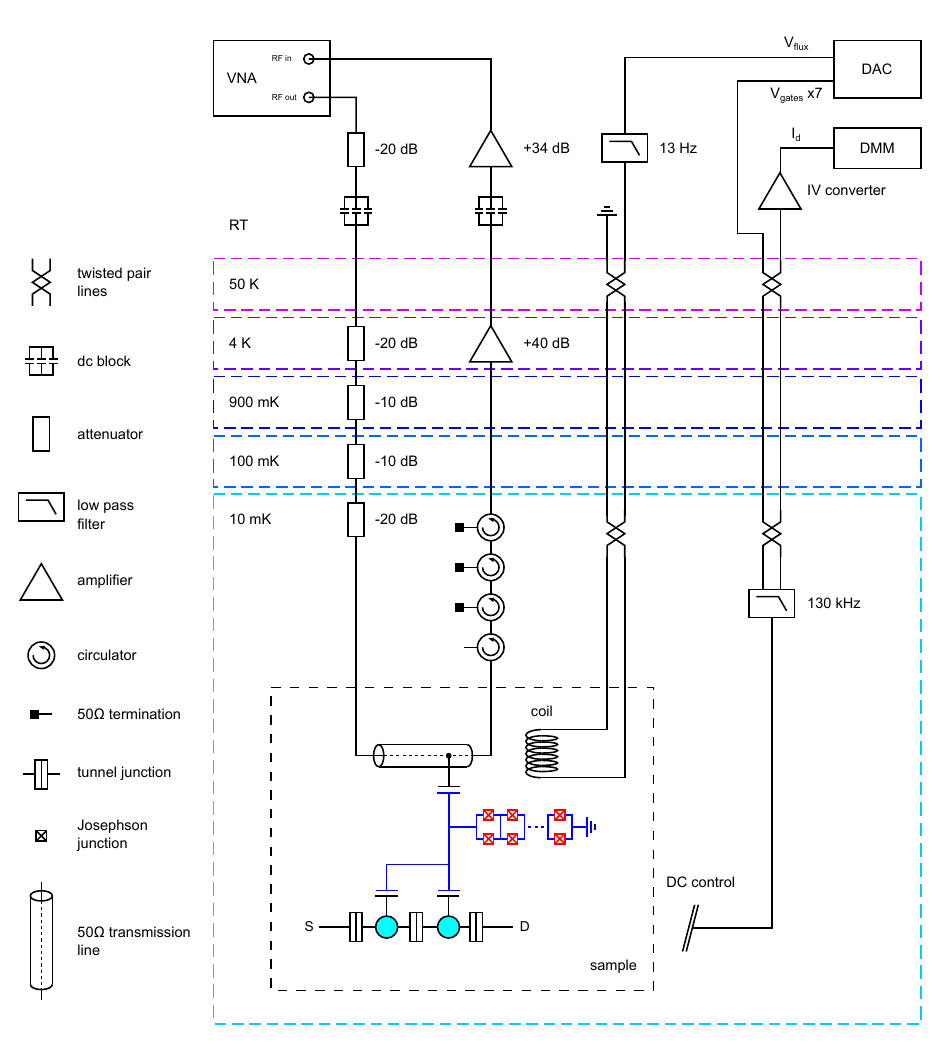}
     \caption{
     \textbf{Schematic of cryogenic and room temperature measurement setup.}
     }
     \label{fig:SupplementarySetup}
\end{figure}

The measurements reported in this work are performed in a dilution refrigerator (Bluefors LD250) at base temperatures around 10 mK (see Supplementary Figure~\ref{fig:SupplementarySetup}).
The device is mounted on a printed circuit board (PCB, QDevil QBoard), which consists of a motherboard with an RC low pass filtering stage (130~kHz cut-off) and a daughterboard that hosts the device.
Good contact between daughter- and motherboard is ensured by spring contacts. The coaxial cables for RF signals are connected directly to the daughterboard.

Gate and bias voltages for the QDs are generated by a 24-channel digital-to-analog converter (DAC, QDevil QDAC-II) and are passed to the sample via twisted pair cables made of phosphor bronze and the RC filtering stage (65~kHz cut-off) on the QBoard.
The drain current through the device is measured by a digital multimeter (DMM, Keysight 34465A) after being amplified by an IV converter (Basel Precision Instruments SP 983C).
The magnetic flux of the SQUID array resonator is controlled by a small superconducting coil mounted directly above the sample.
The coil bias voltage is generated by a DAC (QDevil QDAC-II) and passes through an RC low pass filter (13~Hz cut-off) at room temperature with a total resistance of 1 k$\Omega$ followed by superconducting twisted pair lines that are connected to the coil.

Resonator spectroscopy is performed with a vector network analyzer (VNA, Rohde \& Schwarz ZNB20). The VNA output is attenuated at room temperature, followed by a dc block (Inmet 8039 inner-outer).
The signal passes through an attenuation chain before reaching the 50 $\Omega$ coplanar waveguide (CPW) transmission line on the device.
The transmitted signal then passes through a chain of two circulators (Low Noise Factory CICIC4\_8A) and two isolators (Low Noise Factory ISISC4\_8A). The thrid port of the second circulator is terminated, effectively acting as another isolator. The signal is amplified by a HEMT at the 4 K stage (Low Noise Factory  LNC4\_8C) and by a low noise amplifier at room temperature (Agile AMT\_A0284) after passing through another dc block (Inmet 8039).

\begin{figure}[H]
     \centering
     \includegraphics[width=100mm]{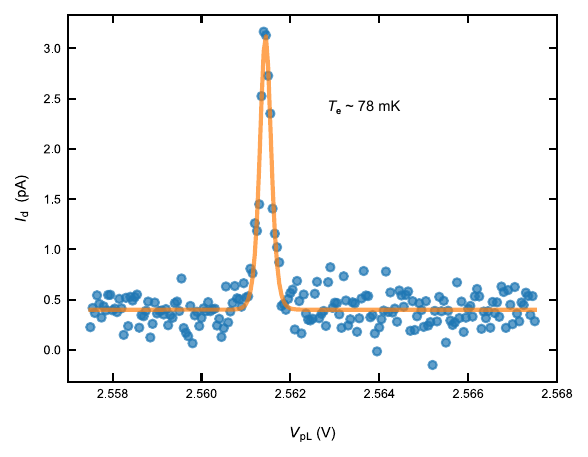}
     \caption{
     \textbf{Electron temperature measurement via Coulomb oscillation.}
     Charge transition line of the left QD recorded with dc-current measurement. Numerical fit to a model (solid orange line, Supplementary Eq.~\ref{eq:e_temp}) results in electron temperature $T_e \sim$ 78 mK.
     }
     \label{fig:SupplementaryElectronicTemp}
\end{figure}

The electron temperature is extracted by measuring Coulomb oscillations. 
When the dot-reservoir tunneling rate $\Gamma_\mathrm{r}$ is low enough, such that $\Gamma_\mathrm{r}<4k_\mathrm{B}T_\mathrm{e}$,
where $k_\mathrm{B}$ is the Boltzmann constant and $T_\mathrm{e}$ is the temperature of the reservoirs, the line shape of the Coulomb peak can be approximated by \cite{s1998_mar}
\begin{equation}
    G(T_\mathrm{e}) = \frac{G_\mathrm{max}}{\cosh\left( \frac{\alpha (V_\mathrm{g} - V_0)}{2k_\mathrm{B} T_\mathrm{e}} \right)^2} + c,
    \label{eq:e_temp}
\end{equation}
where $\alpha = 80$~meV/V is the gate lever arm, $V_\mathrm{g}$ is the gate voltage, $V_0=E_\mathrm{F}/\alpha$ is the gate voltage where the dot level aligns with the Fermi energy of the leads and $c$ is a constant offset.
Fitting the measured current $I_\mathrm{d}$ to $V_\mathrm{bias} G(T_\mathrm{e})$ yields an effective electron temperature of $T_\mathrm{e} \sim$ 78 mK (see Supplementary Figure~\ref{fig:SupplementaryElectronicTemp}).

\newpage
\section{Stability diagrams and number of charges estimation}\label{supplsec2}

\add{Due to the relatively large size of our dots,
they cannot be depleted to the single hole regime within a practical gate voltage range. 
As a consequence, getting a precise particle number is not straightforward.
In this regard, we first present an upper bound of this number based on the saturation carrier density measured when the channel is fully conductive, $p_0$ = 7.54 $\cdot\ 10^{11}$ $\mathrm{cm^{-2}}$.
Based on the QD anisotropy $\sim 0.8$ (see Supplementary Note~\ref{sec:wigner}) and QD radius $l_\mathrm{QD}\sim 70$ nm (extracted from Coulomb diamond measurements, see Supplementary Note~\ref{sec:wigner}), we obtain a QD area $\sim 12000 ~ \mathrm{nm}^2$, which results in a maximum charge number of $\sim 90$.
We note, however, that the percolation density of the heterostructure may provide more precise estimate of the charge number within, as the QDs are formed when the channel is not fully conductive. 
While the exact percolation density of our material is not measured, we roughly estimate the percolation density to be $\sim 0.1p_0 - 0.3p_0$ from the previous literature \cite{s2021_lod, sGe_meff, s2009_tra, s2017_kim, s2015_mi},
which results in a more realistic estimate of the charge number $\sim 9 - 27$.}

\begin{figure}[H]
     \centering
     \includegraphics[width=\textwidth]{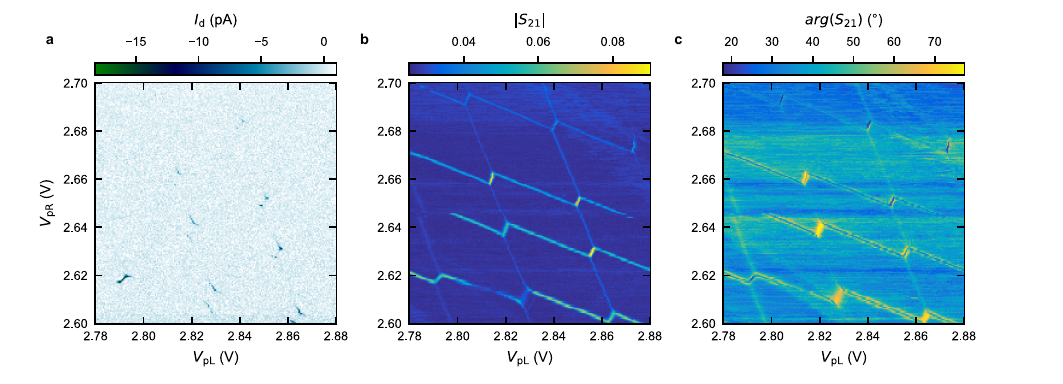}
     \caption{
     \textbf{Zoom-out charge stability diagram.}
     The DQD charge stability diagram, showing the canonical honeycomb pattern \cite{s2003_DQD_rev}, is measured by dc transport (\textbf{a}) through the DQD, as well as detecting magnitude (\textbf{b}) and phase (\textbf{c}) of the feedline transmission $S_{21}$ (at $f_\mathrm{d}=f_\mathrm{r}=5.01$~GHz), as a function of the applied plunger gate voltages $V_\mathrm{pL}$ and $V_\mathrm{pR}$.
     }
     \label{fig:Extended data 2}
\end{figure}

\begin{figure}[H]
     \centering
     \includegraphics[width=0.89\textwidth]{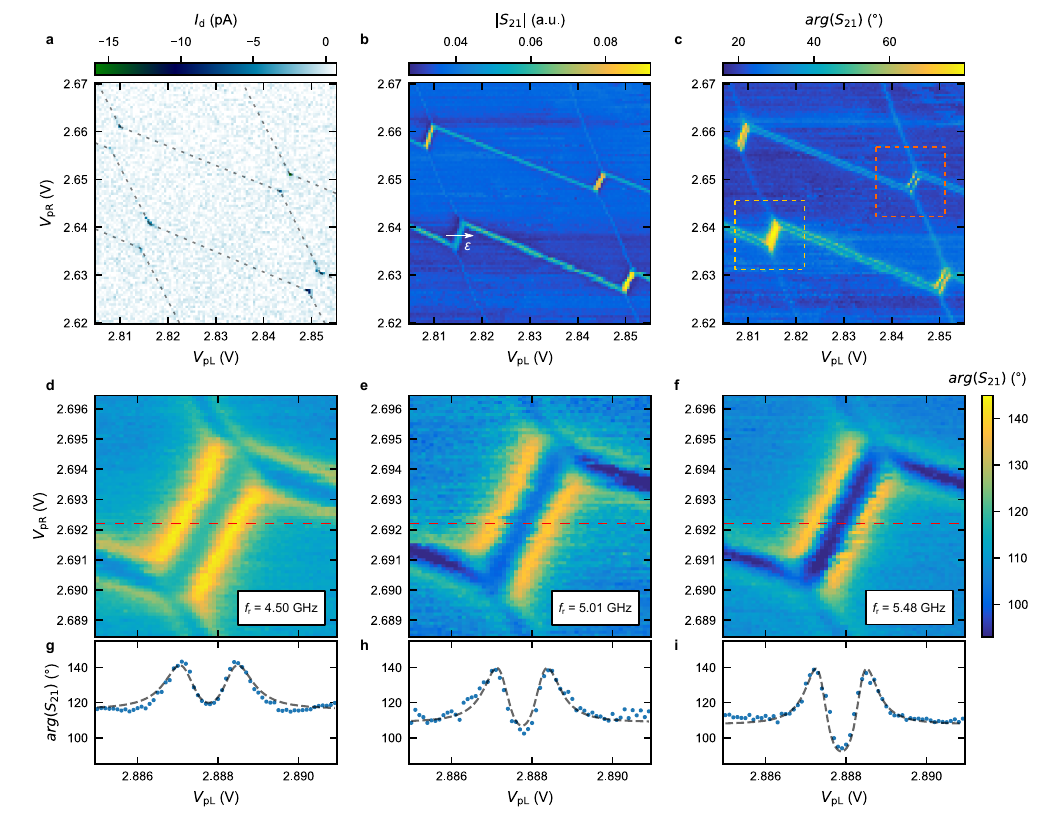}
     \caption{\add{
     \textbf{Extension to Fig. 2 of the main text.} 
     A region of the DQD charge stability diagram as a function of the applied plunger gate voltages $V_\text{pR}$ and $V_\text{pL}$, recorded by dc-transport (\textbf{a}) and by measuring amplitude (\textbf{b}) and phase (\textbf{c}) of the feedline transmission $S_{21}$ at $f_\text{d} = f_\text{r} = 5.01$~GHz. The resonator detects inter-dot and reservoir-dot transitions when 
     their tunneling rates are close to $f_\text{r}$ \cite{s2012_fre}.
     Yellow (orange) dashed box in \textbf{c}: 
     The phase signal increases (decreases) near the inter-dot region with respect to the background, if the resonator is dispersively shifted to lower (higher) frequency. Notably, because the resonator gate lever arm is larger for the right QD, the resonator is more sensitive to its QD-reservoir transitions with respect to those of the left QD.
     \textbf{d-f,} Same inter-dot transition probed with $f_\text{r} = 4.50 ~\mathrm{GHz} < 2t_\text{c}/h$ (\textbf{d}), $f_\text{r} = 5.01 ~\mathrm{GHz} \sim 2t_\text{c}/h$ (\textbf{e}) and $f_\text{r} = 5.48 ~\mathrm{GHz} > 2t_\text{c}/h$ (\textbf{f}).
     The corresponding line-cuts, taken along the red dashed lines, are shown in \textbf{g}, \textbf{h}, and \textbf{i} respectively. The black dashed curves show the simultaneous fit to a master equation (see Methods). Panels \textbf{a}, \textbf{c}, \textbf{e} and \textbf{h} correspond to Figs. 2a - d in the main text, respectively.
     }}
     \label{fig:suppFig2}
\end{figure}

\newpage

\section{Cavity characterization}\label{sec:cav_char}

The superconducting cavity consists of a quarter-wave high-impedance SQUID array resonator. On one side, it is coupled to a 50~$\Omega$ waveguide via a coupling capacitance of $C_\mathrm{ext} \sim 4$~fF and galvanically connected to the Al ground plane on the other side. The resonator has an equivalent lumped capacitance to ground of $C_\mathrm{gnd} = 17$~fF \cite{s2008_Goppl}. The total lumped equivalent capacitance of the resonator is $C_\mathrm{r}\sim 24$~fF, taking into account the parasitic capacitance of the gate lines, estimated to be $\sim 3$~fF from static simulations. The inductance of a single SQUID, extracted from room temperature resistance measurements \cite{s1963_amb}, is $L_\mathrm{SQUID} \sim 0.63$~nH/SQUID. The total lumped equivalent inductance is $L_\mathrm{r} \sim 16$~nH at zero flux. The resulting bare resonance frequency is $f_\mathrm{r} \sim 8.0$~GHz. However, the resonator is operated only from 4 to 6~GHz in the experiments reported in this work. In this frequency range, the resonator lumped equivalent impedance $Z_\mathrm{r} = \sqrt{L_\mathrm{r}/C_\mathrm{r}}$ ranges from 1.6~k$\Omega$ (at $f_\mathrm{r} = 4$~GHz) to 1.1~k$\Omega$ (at $f_\mathrm{r} = 6$~GHz) and the external coupling rate $\kappa_\text{ext}/2\pi$ from 8 MHz to 80 MHz, probably due to the presence of strong standing waves coupled to the resonator around (and above) 6~GHz.
The internal loss rate of the resonator $\kappa_\text{int}/2\pi$ decreases, below 5~GHz and at low photon numbers ($ n_\mathrm{avg} \ll$ 1), to approximately 9~MHz, allowing the resonator to operate in the overcoupled regime in the majority of the explored frequency range \cite{s2021_bla}.
At $f_\mathrm{r} = 5.109$~GHz, roughly in the centre of the frequency operation window, a fit to the master equation of the bare normalized complex feedline transmission, with the charge qubit far detuned, (see Supplementary Eq.~\eqref{eq:S21_bare_resonator} in Supplementary Note~\ref{sec:io_cq}), gives $\kappa/2\pi = 39$~MHz, $\kappa_\mathrm{ext}/2\pi = 29$~MHz, $\kappa_\mathrm{int}/2\pi = 10$~MHz and $n_\mathrm{avg} \sim 0.1$, for a drive power at the resonator input $P_\mathrm{in} \sim-136 $~dBm. The plots of the resulting normalized amplitude $|A/A_0|^2$, phase $arg(A/A_0)$ and complex circle, with the aforementioned fitted parameters, are reported in Supplementary Figure~\ref{fig:cavity_characterization}, where the orange lines \replace{are fitted on top}{represent the fit} of the experimental data (blue dots).
\add{A more detailed study of the evolution of the external and total coupling strengths $\kappa_\mathrm{ext}$ and $\kappa$ as a function of $f_\mathrm{r}$ can be found in Supplementary Figure~\ref{fig:Suppl_kappas_vs_fr}, where the same fit was performed for different flux voltages.}

\begin{figure}[H]
     \centering
     \includegraphics[width=152mm]{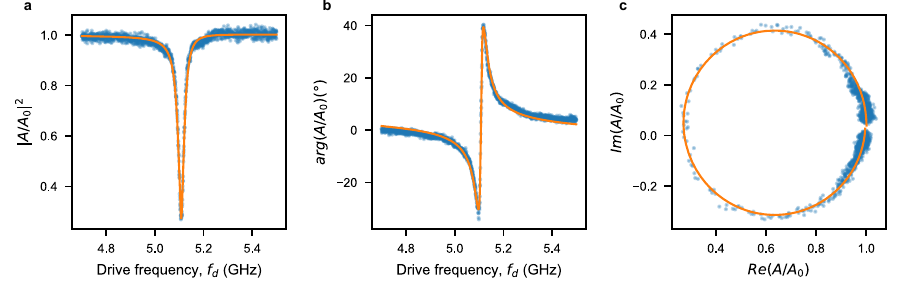}
     \caption{
     \textbf{Bare cavity characterization}. Normalized amplitude $|A/A_0|^2$ (\textbf{a}), phase $arg(A/A_0)$ (\textbf{b}) and complex circle (\textbf{c}) of the feedline transmission as a function of drive frequency $f_\mathrm{d}$ with the charge qubit far detuned from the resonator. Blue dots are experimental data, whereas the orange lines are fits to the master equation.
     }
     \label{fig:cavity_characterization}
\end{figure}

\begin{figure}[H]
     \centering
     \includegraphics[width=120mm]{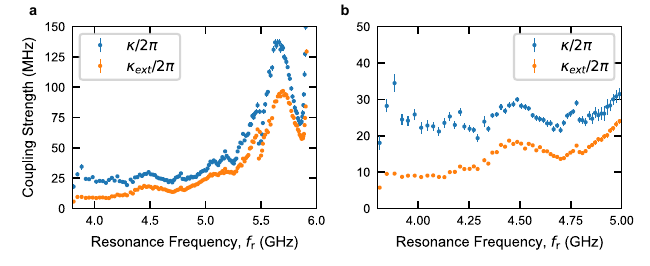}
     \caption{\add{
     \textbf{Resonator coupling strengths as a function of $f_\mathrm{r}$.}
     \textbf{a,} Loaded ($\kappa$) and external ($\kappa_\mathrm{ext}$) coupling strengths of the bare resonator as function of $f_\mathrm{r}$ extracted from fitting frequency line-cuts of a flux sweep measurement.
     The large increase of $\kappa_\mathrm{ext}$ towards 6~GHz is attributed to coupling to a standing wave.
     \textbf{b,} Zoom-in of \textbf{a} to the most interesting frequency range for the presented experiments.
     }}
     \label{fig:Suppl_kappas_vs_fr}
\end{figure}

\add{Some of the spectroscopies reported in this work show some extra narrow transitions (Fig. 4a top panel, Fig. 5a top panel and Fig. 6a third panel from the top). To clarify the nature of these extra transitions, we acquire the resonator response as a function of the magnetic flux with all the gate lines grounded, i.e. with no charge qubit defined. The result is reported in Supplementary Figure~\ref{fig:Suppl3_Spurious_modes}a. With a careful look at the high-resolution spectrum, many avoided crossings can be observed, as clearly indicated by the orange arrows in the zoom-in reported in Supplementary Figure~\ref{fig:Suppl3_Spurious_modes}b. We attribute them to the coupling of the high-impedance resonator to two-level systems (TLSs) located in the tunneling junctions \cite{s2019_Mul}. A high-power magnetic flux sweep, not reported here, shows the disappearance of the avoided crossings, indicating a saturation of the TLSs and further supporting our hypothesis. We note that these TLS modes may interact with the qubit mediated by the resonator, which may result in a weak DQD detuning dependence in the spectrum, such as the ones observed in Figs. 5a (top panel) and 6a (third panel from top) in the main manuscript.
}

\add{
The measurements shown in Supplementary Figures~\ref{fig:Suppl3_Spurious_modes}a and b, however, were acquired in a new cooldown and for this reason some of these avoided crossings appear at different frequencies with respect to the reported measurements. We thus report the same flux sweep in Supplementary Figure~\ref{fig:Suppl3_Spurious_modes}c, together with the corresponding zoomed-in curve (orange box) in Supplementary Figure~\ref{fig:Suppl3_Spurious_modes}d, but taken in the same cooldown of the experiments reported in our work. Here, a charge qubit is hybridizing with the resonator at $\sim5.2$~GHz to result in vacuum-Rabi mode splitting (red dashed line in Supplementary Figures~\ref{fig:Suppl3_Spurious_modes}c and d). Also in this case, several avoided crossings are visible, both above and below the charge qubit frequency. This strongly suggests that these extra transitions cannot be attributed to extra states in the DQD, but most likely are due to TLSs in the SQUID array resonator.
In the zoom-in of Supplementary Figure~\ref{fig:Suppl3_Spurious_modes}d, three avoided crossings are clearly visible. The two around 5~GHz correspond to those observed in Fig.~6a (third panel from the top). The highest frequency avoided crossing matches the one visible in Fig.~4a (top panel).
}

\begin{figure}[H]
     \centering
     \includegraphics[width=152mm]{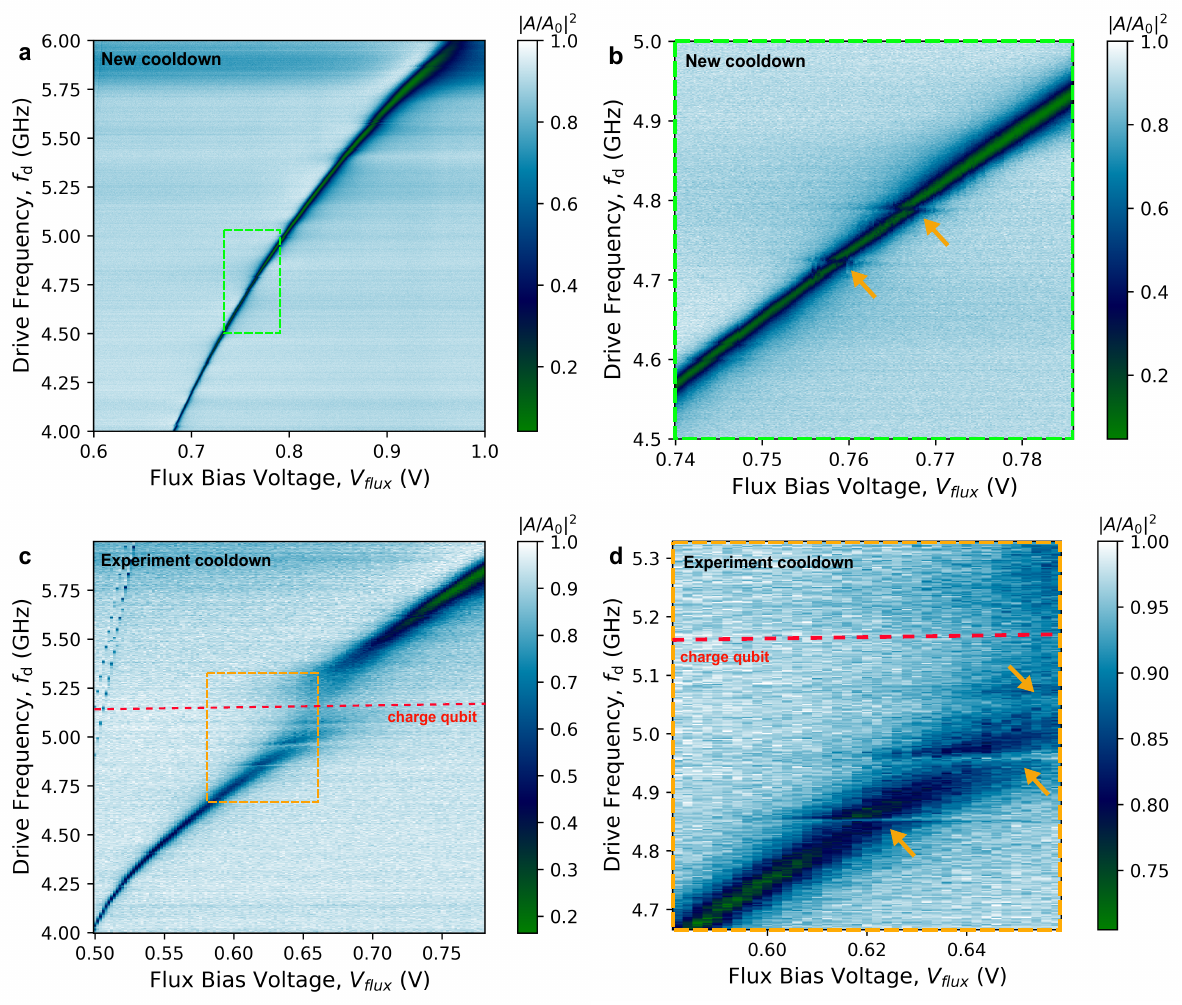}
     \caption{\add{
     \textbf{Observation of the spurious two-level systems in the SQUID array resonator. a,} Normalized feedline transmission $|A/A_0|^2$ as a function of the resonator drive frequency $f_\mathrm{d}$ and flux voltage $V_\mathrm{flux}$ with a charge qubit far detuned from the resonator. The measurement is taken with a low photon number ($n_\mathrm{ph}\ll 1$). 
     \textbf{b,} Zoom-in of the green-box region in \textbf{a}. Orange arrows denote the signatures of two-level systems (TLSs) interacting with the resonator.
     \textbf{c,} $|A/A_0|^2$ as a function of $f_\mathrm{d}$ and $V_\mathrm{flux}$ with the DQD charge qubit in resonance with the resonator at $\sim 5.2$~GHz (red dashed line).
     \textbf{d} Zoom-in of the orange box region in \textbf{c}. Orange dashed lines denote the signatures of TLSs, unrelated to the DQD charge qubit.
     \textbf{c} is taken in the same cooldown together with the datasets presented in the manuscript, while \textbf{a} is taken in a separate cooldown.
     }}
     \label{fig:Suppl3_Spurious_modes}
\end{figure}

\newpage

\section{Input-output theory for charge qubits}\label{sec:io_cq}
The analytic derivation of the resonator response follows the standard procedure of combining the Heisenberg-Langevin equation of motion for the cavity field operator \textit{a} with the input-output relation \cite{sCollett_1984}. For a bare hanger resonator, in a frame rotating with the driving frequency, the equation of motion reads \cite{sIO_hanged}:
\begin{equation}
    \dot a = -i\Delta_\mathrm{r} a - \frac{\kappa_\mathrm{ext}+\kappa_\mathrm{int}}{2}a -\sqrt{\frac{\kappa_\mathrm{ext}}{2}}b_\mathrm{in},
\end{equation}
where $\Delta_\mathrm{r}$ is the resonator-drive detuning $\omega_\mathrm{r} - \omega_\mathrm{d}$, $\kappa_\mathrm{ext}$ is the total coupling strength of the cavity to the \textit{b} modes of the waveguide, $\kappa_\mathrm{int}$ is the internal resonator dissipation and $b_\mathrm{in}$ is the input field. The steady-state solution of the cavity field ($\dot a = 0$) is easily found:
\begin{equation}
    a = \frac{-\sqrt{\kappa_\mathrm{ext}/2}}{i\Delta_\mathrm{r}+\kappa/2}b_\mathrm{in},     
\end{equation}
with $\kappa = \kappa_\mathrm{ext} + \kappa_\mathrm{int}$, and combined with the input-output relation: $b_\mathrm{out}=b_\mathrm{in}+\sqrt{\frac{\kappa_\mathrm{ext}}{2}}a$ \cite{sCollett_1984} to get the scattering coefficient:
\begin{equation}
    S_\mathrm{21} = \frac{b_\mathrm{out}}{b_\mathrm{in}} = 1 + \frac{\kappa_\mathrm{ext}/2}{i\Delta_\mathrm{r}+\kappa/2} = \frac{\Delta_\mathrm{r}-i\kappa_\mathrm{int}/2}{\Delta_\mathrm{r}-i\kappa/2}.
    \label{S21_bare}
\end{equation}

The measured resonator response, however, can deviate from the ideal case because of environmental factors and/or impedance mismatches in proximity of the resonator or between input and output ports \cite{sProbst_2015}. The environment is usually modeled with the complex term:
\begin{equation}
    ae^{i\alpha}e^{-2\pi if_\mathrm{d}\tau},
\end{equation}
where $a$ ($\alpha$) is a rescaling amplitude (phase shift), $\tau$ is the electric delay due to the cable length and $f_\mathrm{d}$ the drive frequency. Impedance mismatches are taken into account by a complex $\kappa_\mathrm{ext}=|\kappa_\mathrm{ext}|e^{i\phi}$, resulting in the corrected scattering coefficient:
\begin{equation}
    S_\mathrm{21} = ae^{i\alpha}e^{-2\pi if_\mathrm{d}\tau} \ \frac{\Delta_\mathrm{r}-i(\kappa-|\kappa_\mathrm{ext}|e^{i\phi})/2}{\Delta_\mathrm{r}-i\kappa/2}.
    \label{eq:S21_bare_resonator}
\end{equation}

In the aforementioned experiments, a charge qubit is coupled to the superconducting resonator. The charge qubit Hamiltonian can be written as $H_\mathrm{cq} = \frac{\varepsilon}{2}\sigma_\mathrm{z} + t_\mathrm{c}\sigma_\mathrm{x}$, with the DQD detuning $\varepsilon$, the inter-dot tunneling coupling $t_\mathrm{c}$ and the Pauli operators $\sigma_x$ and $\sigma_z$.
The general resonator-charge qubit Hamiltonian is expressed as
\begin{equation}
    H/\hbar=\omega_\mathrm{r}a^\dag a+\frac{\omega_\mathrm{q}}{2}\sigma_z+g_\mathrm{eff}\sigma_x(a+a^\dag),
\end{equation}
where $\omega_\mathrm{r}/2\pi$ ($\omega_\mathrm{q}/2\pi=\sqrt{\varepsilon^2+4t_\mathrm{c}^2}/h$) is the resonator (qubit) frequency, $g_\mathrm{eff}=g_0\frac{2t_\mathrm{c}}{\hbar\omega_\mathrm{q}}$ is the effective charge-photon coupling strength with $g_0$ representing the coupling strength at $\varepsilon=0$ and $a$ ($a^\dag$) is the photon annihilation (creation) operator. Here, we focus only on transversal interactions $\sigma_x (a + a^\dag)$, i.e. through the DQD detuning degree of freedom $\varepsilon$, because purely longitudinal interactions ($\sigma_z (a + a^\dag)$) do not leave spectroscopic signatures. A unitary transformation $U=\exp[-i\omega_\mathrm{d}t (a^\dag a + \sigma_z/2)]$ allows us to rewrite the Hamiltonian in the drive frame ($\omega_\mathrm{d}$), using the rotating wave approximation to neglect fast rotating terms:
\begin{equation}
    H_\mathrm{RWA}=\Delta_\mathrm{r}a^\dag a+\frac{\Delta_\mathrm{q}}{2}\sigma_z+g_\mathrm{eff}(a^\dag \sigma_\mathrm{-} + a\sigma_\mathrm{+}),
\end{equation}
with $\Delta_\mathrm{q}$ being the qubit-drive detuning $\omega_\mathrm{q}-\omega_\mathrm{d}$ and $\sigma_\mathrm{+}$ and $\sigma_\mathrm{-}$ the qubit raising and lowering operators. The result is the well-known Jaynes-Cummings Hamiltonian. Due to the qubit-resonator interaction, the Heisenberg-Langevin equation of motion for the cavity field $a$ shows an extra term ($[a, g_\mathrm{eff}(a^\dag \sigma_\mathrm{-}+a\sigma_\mathrm{+})]$):
\begin{equation}
    \dot a = -i\Delta_\mathrm{r} a - \frac{\kappa_\mathrm{ext}+\kappa_\mathrm{int}}{2}a -\sqrt{\frac{\kappa_\mathrm{ext}}{2}}b_\mathrm{in}-ig_\mathrm{eff}\sigma_\mathrm{-}.
    \label{eq:a_derivative}
\end{equation}

The steady-state solution for the cavity field requires the knowledge of $\sigma_\mathrm{-}$. The time evolution of the qubit lowering operator is again captured by the Heisenberg picture:
\begin{equation}
    \dot \sigma_\mathrm{-} = -i\Delta_\mathrm{q} \sigma_\mathrm{-} - \Gamma\sigma_\mathrm{-} - ig_\mathrm{eff}\sigma_za,
    \label{eq:low_der}
\end{equation}
with $\Gamma$ the qubit decoherence rate (including both relaxation and dephasing).
Assuming the qubit to be in a thermal state, the expectation value of $\sigma_z$ can be
expressed as the average probability for the qubit to be in the ground ($p_\mathrm{0}$) or in the excited ($p_\mathrm{1}$) state as:
\begin{equation}
    \langle\sigma_\mathrm{z}\rangle = p_\mathrm{0}-p_\mathrm{1} = \tanh(\hbar \omega_\mathrm{q}/k_\mathrm{B}T),
     \label{eq:exp_val_sigmaz}
\end{equation}
where $k_\mathrm{B}$ is the Boltzmann constant and $T$ the ``experimental" temperature, i.e. the temperature of any kind of bath (phonons, traps, reservoirs, etc.) that can exchange thermal energy with the qubit.
In the experiment ($\omega_\mathrm{q} > 3$~GHz), for $T \approx 10$ mK, $\langle\sigma_\mathrm{z}\rangle \approx p_\mathrm{0}$, i.e. thermal excitations can be neglected and the qubit is always in the ground state when not driven.
Inserting Supplementary Eq.~\eqref{eq:exp_val_sigmaz} into Supplementary Eq.~\eqref{eq:low_der} and setting $\dot\sigma_\mathrm{-} = 0$ results in the expectation value of the qubit lowering operator:
\begin{equation}
    \sigma_\mathrm{-} = \frac{g_\mathrm{eff}a}{-\Delta_\mathrm{q}+i\Gamma}.
    \label{eq:SE_sigma-}
\end{equation}

Inserting Supplementary Eq.~\eqref{eq:SE_sigma-} into Supplementary Eq.~\eqref{eq:a_derivative} and setting $\dot a = 0$, the cavity field in the hybridized case is obtained:
\begin{equation}
    a = \frac{-\sqrt{\kappa_\mathrm{ext}/2}}{i\Delta_\mathrm{r}+\kappa/2+\frac{ig_\mathrm{eff}^2}{-\Delta_\mathrm{q}+i\Gamma}}b_\mathrm{in}.
    \label{eq:SE_a}
\end{equation}

Finally, introducing the DQD susceptibility $\chi=\frac{\sigma_\mathrm{-}}{a}=\frac{g_\mathrm{eff}}{-\Delta_\mathrm{q}+i\Gamma}$ and following the same steps of Supplementary Eq.~\eqref{S21_bare}, the expression of $S_\mathrm{21}$ for the hybridized system is found, including the correction factors for the environment:
\begin{equation}\label{eq:S21_coupled}
    S_\mathrm{21} = ae^{i\alpha}e^{-2\pi if_\mathrm{d}\tau} \ \frac{\Delta_\mathrm{r}-i(\kappa-|\kappa_\mathrm{ext}|e^{i\phi})/2+g_\mathrm{eff}\chi}{\Delta_\mathrm{r}-i\kappa/2+g_\mathrm{eff}\chi}.
\end{equation}

\newpage
\section{Charge qubit spectroscopies}\label{sec:cq_spectrum}

\begin{figure}[H]
     \centering
     \includegraphics[width=0.5\textwidth]{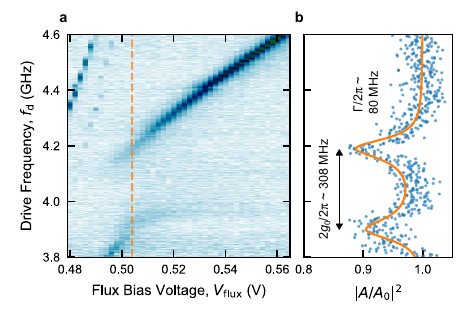}
     \caption{\add{
     \textbf{Extension to Fig. 3 of the main text.}
     \textbf{a,} Normalized amplitude of feedline transmission $|A/A_\mathrm{0}|^2$ as a function of drive frequency $f_\mathrm{d}$ and the voltage $V_\text{flux}$ applied to the resonator coil which tunes the resonator frequency $f_\text{r}$. During the measurement, the DQD is kept at $\varepsilon = 0$.
     An avoided crossing is observed around $V_\text{flux} = 504$ mV, when the bare resonator frequency $f_\text{r}$ matches the DQD charge transition ($f_r=f_q=2t_\text{c}/h$).
     \textbf{b,} Frequency line-cut at the avoided crossing along the orange dashed line in \textbf{a}.
     A fit to the master equation model is represented by a solid orange line (see Methods). The extracted values for $2g_0$ and $\Gamma$ are reported in Supplementary Table~\ref{table:ext_table_chargequbit} (2$t_\text{c}/h = 4.036$~GHz).
     }}
     \label{fig:fig3_v5_flux}
\end{figure}

\begin{figure}[H]
     \centering
     \includegraphics[width=0.9\textwidth]{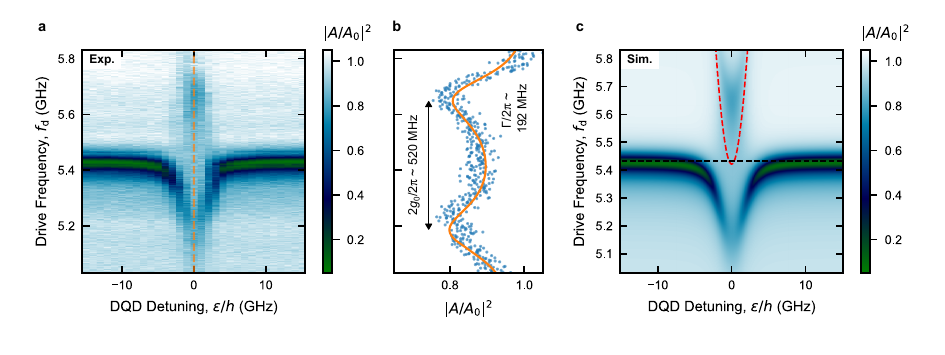}
     \caption{
     \textbf{Strong charge-photon coupling at the charge sweet spot.}
     \textbf{a,} Normalized amplitude of feedline transmission $|A/A_\mathrm{0}|^2$ as a function of drive frequency $f_\mathrm{d}$ and DQD detuning $\varepsilon$. An avoided crossing - the signature of the strong coupling regime - is observed when the DQD-charge transition matches the bare resonator frequency.
     \textbf{b,} Frequency line-cut (along the orange dashed line in \textbf{a}) at resonance, highlighting the vacuum-Rabi splitting $2g_0/2\pi$.
     A numerical fit to the master equation model is represented by a solid orange line. The extracted values for $2g_0$ and $\Gamma$ are indicated (2$t_\text{c}/h = 5.433$~GHz) and reported in Supplementary Table~\ref{table:ext_table_chargequbit}.
     \textbf{c,} Simulation of $|A/A_0|^2$ using input-output theory with the parameters extracted (see Supplementary Table 1) by fitting the dataset in panel \textbf{b} to the master equation model (see Methods). 
     }
     \label{fig:ext_fig2}
\end{figure}

\begin{table}[h!]
\centering
\begin{tabularx}{0.8\linewidth}{p{2.5cm}XXX}\toprule
Parameter & Supplementary Figure~\ref{fig:fig3_v5_flux}b & Fig.~3c & Supplementary Figure~\ref{fig:ext_fig2}b\\
\midrule
$t_\mathrm{c}/h$ (GHz) & 2.018 & 2.072 & 2.712 \\
$f_\mathrm{r}$ (GHz) & 4.052 & 4.149 & 5.432$^*$ \\
$g_0/2\pi$ (MHz) & 154 & 165 & 260 \\
$\Gamma/2\pi$ (MHz) & 80 & 57 & 192 \\
$\kappa^*/2\pi$ (MHz) & 23 & 19 & 61 \\
$\kappa^*_\mathrm{ext}/2\pi$ (MHz) & 9 & 8 & 42 \\
$C$ & 52 & 100 & 23 \\
\bottomrule
\end{tabularx}
\caption{Extracted parameters from the three hybrid system configurations shown in Fig.~3. The parameters denoted with $^*$ are obtained from a separate measurement of the bare resonator and used as fixed parameters in the master equation fit.
The discrepancy between the parameters extracted by $V_\mathrm{flux}$ (Fig.~3b) and sweeping $\varepsilon$ (Figs.~3c) is attributed to a jump of the charge qubit to a slightly higher \replace{$t_\mathrm{c}$}{qubit frequency} and to the shorter integration time used for the latter measurement.}
\label{table:ext_table_chargequbit}
\end{table}

\begin{figure}[H]
     \centering
     \includegraphics[width=90mm]{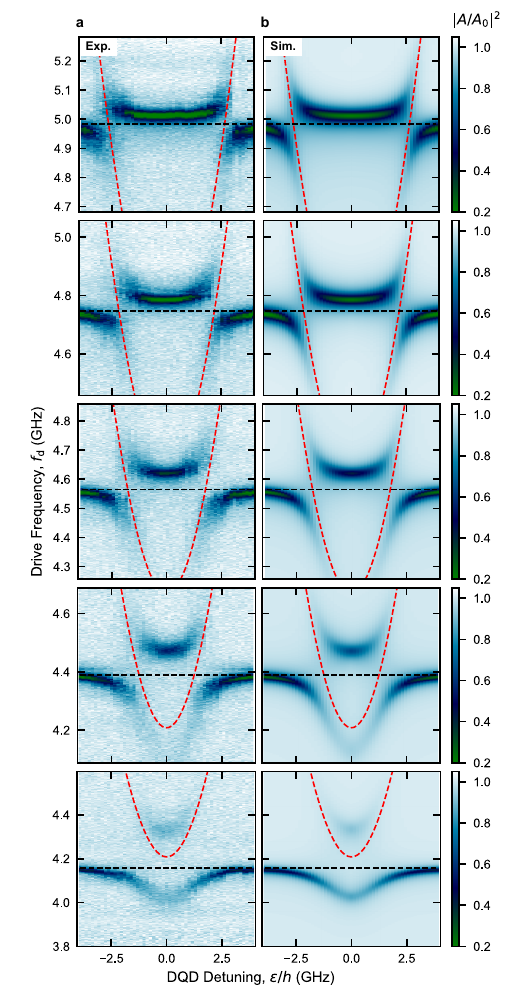}
     \caption{
     \textbf{Detailed spectroscopy of the charge qubit.}
     \textbf{a,} Normalized amplitude of feedline transmission $|A/A_\mathrm{0}|^2$ as a function of drive frequency $f_\mathrm{d}$ and DQD detuning $\varepsilon$ for five different resonator frequencies (for a constant DQD $t_c$), revealing the charge qubit dispersion relation. The black (red) dashed line represents bare resonator (DQD-charge qubit excitation) frequency, obtained using the extracted fitting parameters.
     Some of these datasets are already presented in Fig.~4 \add{in the main text}.
     \textbf{b,} Simulation of the normalized amplitude of feedline transmission $|A/A_\mathrm{0}|^2$ using input-output theory with the parameters reported in \replace{Extended Data }{Supplementary} Table~\ref{table:Extended data Table 2}. 
     }
     \label{fig:Extended data 3}
\end{figure}

\begin{table}[h!]
\centering
\begin{tabularx}{0.7\linewidth}{p{3cm}XXXXX}\toprule
Parameter & Panel 1 & Panel 2 & Panel 3 & Panel 4 & Panel 5 \\
\midrule
$f_\mathrm{r}$ (GHz)               & 4.156	& 4.389	& 4.565	& 4.747	& 4.983	\\
$g_0/2\pi$ (MHz)                   & 155	& 147	& 150	& 148	& 151	\\
$\kappa*/2\pi$ (MHz)               & 19	    & 20	& 22	& 32	& 37	\\
$\kappa_\mathrm{ext}*/2\pi$ (MHz)  & 8	    & 11	& 13	& 21	& 22	\\
\bottomrule
\end{tabularx}
\caption{
\replace{Extracted parameters from the five hybrid system configurations shown in Supplementary Figure~\ref{fig:Extended data 3} numbered from bottom to top.
The inter-dot tunnel coupling $t_\mathrm{c}/h = 2.087$~GHz is fitted only for the first panel and fixed to the extracted value for the remaining panels.}
{
Extracted individual fitting parameters from the five hybrid system configurations shown in Supplementary Figure~\ref{fig:Extended data 3} numbered from bottom to top.
The fitted values for the shared parameters are $t_\mathrm{c}/h = 2.104$~GHz, $\beta_\mathrm{pL} = 23.8$~meV/V, $\Gamma_0 = 57$~MHz, $\Gamma_\varepsilon = 164$~MHz.
}
The parameters denoted with $^*$ are obtained from an independent measurement of the bare resonator and used as fixed parameters in the master equation fit.
}
\label{table:Extended data Table 2}
\end{table}

\newpage
\section{Coulomb interactions in doubly occupied QDs and WM formations}\label{sec:wigner}

\begin{figure}[H]
     \centering
     \includegraphics[width=1\textwidth]{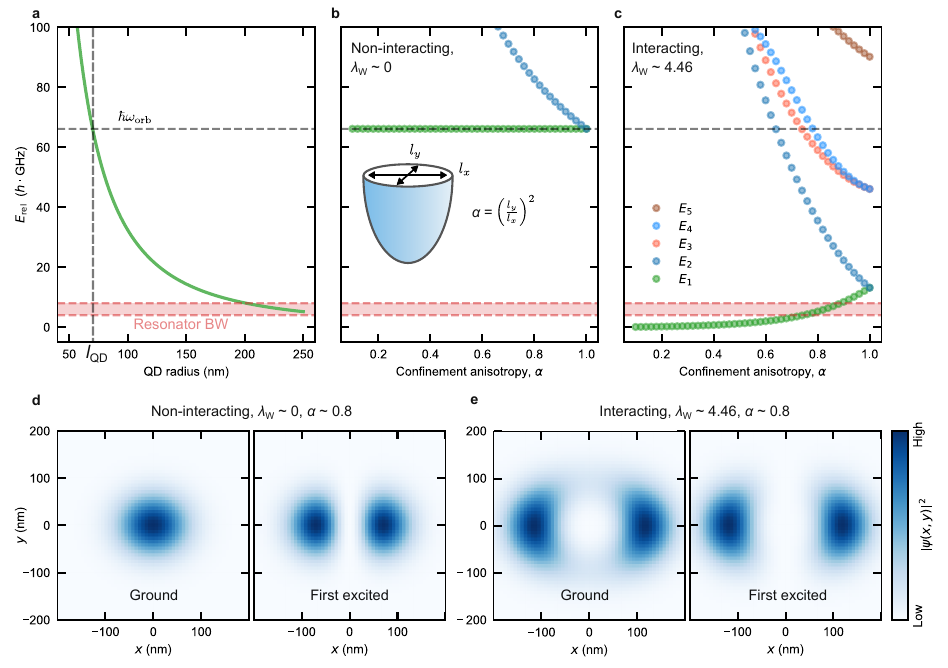}
     \caption{\replace{
     \textbf{Dependence of the exchange-split states on the confinement anisotropy.}
     Calculated eigenenergies of a two-body QD as a function of the confinement anisotropy $\alpha = (l_y/l_x)^2$, with the Wigner ratio $\lambda_\mathrm{w} = E_\mathrm{ee}/\hbar\omega_\mathrm{orb} = 4.46$ \cite{s2021_aba}. Inset: The energy splitting between the lowest two levels i.e. the singlet-triplet splitting $\Delta_\mathrm{ST}$. 
     The dots of different colors denote distinct spectra up to the 8 lowest eigenstates ($E_1 - E_8$) of Supplementary Eq.~\eqref{eq:SE2.2}.
     }{
     \textbf{Orbital state renormalization due to Coulomb interaction and confinement anisotropy.} ~~
     \textbf{a,} Orbital energy splitting $E_1$ between ground and first excited state without Coulomb correlation effect as a function of the hole quantum dot (QD) radius. $l_\mathrm{QD} \sim 70$~nm denotes the radius of the QD in this work, with the corresponding $\omega_\mathrm{orb}/2\pi \sim 70$~GHz. The red shaded region denotes the frequency bandwidth of our resonator $4-8$~GHz. An effective hole mass of $m^* \sim 0.057m_\mathrm{e}$ is assumed, with $m_\mathrm{e}$ being the free electron mass.
     \textbf{b} (\textbf{c}), $E_\mathrm{rel}$ as a function of confinement anisotropy $\alpha$ without (with) finite Coulomb correlation effect. A schematic of the confinement potential is shown in \textbf{b}, where $l_x$ ($l_y$) denotes the characteristic length scale of the confinement along the major (minor) axis with $\alpha = (l_y / l_x)^2 \le 1$. The Wigner ratio $\lambda_\mathrm{W} = E_\mathrm{ee}/\hbar\omega_\mathrm{orb}$ quantifies the electron-electron interaction energy with respect to confinement energy. In \textbf{c}, $\lambda_\mathrm{W} = 4.46$ corresponds to the expected Wigner ratio in our QD.
     $E_1$ (green dots) represents the minimal orbital splitting of the QD corresponding to $l_x > l_y$.
     \textbf{d} (\textbf{e}), Charge density function spanned around the center of a QD for non-interacting (interacting) case with $\alpha \sim 0.8$. 
     Driven by Coulomb correlation, the ground state charge density (left panel in \textbf{e}) becomes similar to that of the excited state (right panel in \textbf{e}).
     }}
     \label{fig:WM_Levels}
\end{figure}

Supplementary Figure~\ref{fig:WM_Levels}a presents the expected orbital splitting (Supplementary Eq.~\eqref{eq:SE2.1}) \add{in planar Ge} as a function of \replace{a quantum dot radius in planar Ge}{the QD radius $l_\mathrm{QD}$}, assuming an effective hole mass of $m_\text{HH}^* = 0.057 m_\mathrm{e}$ \cite{s2021_sca}, where $m_\mathrm{e}$ is the free electron mass.
The geometry of our device supports a QD radius of $l_\mathrm{QD} \sim 70$ nm, from where we estimate the orbital splitting $\hbar \omega_\mathrm{orb}$:
\begin{equation} \label{eq:SE2.1}
\hbar\omega_\mathrm{orb} = \frac{\hbar^2}{m^*l_\mathrm{QD}^2} \sim \; 70 \; h\cdot\mathrm{GHz}.
\end{equation}
\add{The red shaded region in Supplementary Figure~\ref{fig:WM_Levels}a illustrates the resonator frequency bandwidth $4-8$~GHz in our setup which demonstrates that a QD with $l_\mathrm{QD}> 200$~nm is needed to decrease the orbital splitting below 8~$h\cdot$GHz.}


\replace{To justify the assumption of the WM states, we first write the two-body Hamiltonian as Eq. \eqref{eq:SE2.2}, neglecting the possible anisotropies of the effective mass:}
{While more elaborated calculations such as full-configuration-interaction (FCI) calculation may capture the complete picture of the dynamics of our QD with multi particles, 
we provide a simplified two-body Hamiltonian (Supplementary Eq.~\eqref{eq:SE2.2}) also considering Coulomb interaction and confinement anisotropy $\alpha$. For simplicity, we neglect the possible anisotropies of the effective mass in the Hamiltonian:}
\begin{gather} 
H = H_0(x_1, y_1) + H_0(x_2, y_2) + V_\mathrm{int}(x_1-x_2, y_1-y_2) \\
H_0(x, y) = 
\frac{p_x^2 + p_y^2}{2m^*} + 
\frac{m^*\omega_\mathrm{orb}^2}{2}\left(x^2 + \frac{y^2}{\alpha^2}\right) 
\label{eq:SE2.2} \\
V_\mathrm{int}(x, y) = \frac{1}{4\pi\varepsilon}\frac{e^2}{\sqrt{x^2 + y^2}}.
\end{gather}

We consider the case with a QD anisotropy $\alpha = (l_y/l_x)^2 \leq 1$ caused by the confinement potential \cite{s2004_dro}, implying that the QD length $l_y$ along the $y$-axis is shorter than the length $l_x$ along $x$-axis and thus $\omega_\mathrm{orb} \propto 1/l_x^2$. With the electron-electron interaction energy $E_\mathrm{ee} = {e^2}/{4\pi\varepsilon l_x}$ given by the Coulomb interaction between two particles separated by $l_x$, the Wigner ratio $\lambda_\mathrm{w} = E_\mathrm{ee}/\hbar\omega_{orb} \propto l_x$ quantifies the ratio between Coulomb interaction and confinement potential. \replace{Considering $\varepsilon \sim 16$, $m^* = m_\mathrm{HH}^*$ and $l_x \sim 70$ nm, we obtain $\lambda_\mathrm{W} = 4.46$ which justifies the WM states in this work.}
{Here, we note $\lambda_\mathrm{W} \sim 4.46$ is estimated from our QD radius $\sim 70$ nm.}

To study the dependence of the energy scale on $\alpha$, we separate the full Hamiltonian Supplementary Eq.~\eqref{eq:SE2.2} in the center of mass (COM) coordinate $\mathbf{R} = (\mathbf{r}_1 + \mathbf{r}_2)/\sqrt{2}$ and the relative coordinate $\mathbf{r} = (\mathbf{r}_1 - \mathbf{r}_2)/\sqrt{2}$ with $\mathbf{r}_i = (x_i, y_i)$, as well as $x=x_1-x_2$ and $y=y_1-y_2$.
This allows the Hamiltonian to be described by $H = H_\mathbf{R} + H_\mathrm{rel}$ where $H_\mathbf{R}$ is the non-interacting 2D harmonic oscillator Hamiltonian in the COM coordinate \cite{s2021_aba} and $H_\mathrm{rel}$ is the Hamiltonian in the relative coordinate. 
\add{We note that while the method of COM and relative coordinates well describes the dynamics of a harmonic single QD, elaborated models such as two-center-oscillator (TCO) may provide more precise description of the dynamics in a DQD \cite{s2022_yan}.}
By rescaling $l_x$ to $\tilde{x} = x/l_x$ and $\tilde{y} = y/l_x$, we find   
\begin{equation} \label{eq:SE2.3}
\frac{H_\mathrm{rel}}{\hbar\omega_\mathrm{orb}} = 
\frac{\tilde{p}_x^2 + \tilde{p}_y^2}{2} + 
\frac{\tilde{x}^2 + \tilde{y}^2/\alpha^2}{2} + 
\frac{\lambda_\mathrm{W}}{\sqrt{\tilde{x}^2 + \tilde{y}^2}}.
\end{equation}

Because $H_\mathrm{rel}$ can capture the \delete{Wigner molecule
}physics \add{of strongly-correlated states}, we numerically diagonalize Supplementary Eq.~\eqref{eq:SE2.3} to obtain the eigenenergies $E_\mathrm{rel}$ as a function of $\alpha$ in Supplementary Figures~\ref{fig:WM_Levels}b and c. 
\delete{Due to the large $\lambda_\mathrm{W}$, weak anisotropy ($\alpha \sim 1$) suffices to introduce the quenched energy spectrum in line with the observation in this work \cite{s2008_gol}.}
The total energy of the DQD system is then given by $E_\mathrm{tot}= E_\mathrm{rel}+E_\mathrm{R}= E_\mathrm{rel}+ \hbar \omega_\mathrm{orb}m$, with integer $m$.
\add{
Supplementary Figure~\ref{fig:WM_Levels}b (c) demonstrates the orbital splitting as a function of $\alpha$ with $\lambda_\mathrm{W} \sim 0~(4.46)$ which represents the non-interacting (interacting) case resulting from Supplementary Eq.~\eqref{eq:SE2.3}. 
Apparently, finite Coulomb interaction quenches down the singlet-triplet splitting (green dots in Supplementary Figures~\ref{fig:WM_Levels}b and c) from $\sim 70 ~h\cdot$GHz to $\sim 14 ~h\cdot$GHz in the perfectly anisotropic case ($\alpha = 1$).
Additionally for $\alpha < 1$, the singlet-triplet splitting decreases further down in energy only in the interacting case (Supplementary Figure~\ref{fig:WM_Levels}c) where we find $\alpha \sim 0.8$ is required to have the excited orbital state within the resonator bandwidth in this two-body model.
Furthermore, we also present the effect of Coulomb correlation and confinement anisotropy in the shape of the ground and excited state charge density in Supplementary Figures~\ref{fig:WM_Levels}d and e. 
In the interacting case (Supplementary Figure~\ref{fig:WM_Levels}e), the ground state charge density (left panel) becomes similar to that of the excited state (right panel) which is often referred to as a Wigner molecule \cite{s2021_aba, s1999yannouleas, s2009li, s2007_yan, s2022_yan}.
While the Coulomb interaction and QD confinement anisotropy may have different implications in the specific charge density of the multi-hole QD ground state, 
our calculation indicates that the orbital state renormalization driven by Coulomb correlation and QD anisotropy results in the low-lying states observed in Fig. 5 and 6.
}

\section{Input-output theory for strongly-correlated states}\label{sec:io_wigner}

As shown in Figs. 5, 6 and in the Hamiltonian in Methods section, the strongly-correlated states (SCSs) show a multi-level structure. Thereby, a generalization of Supplementary Eq.~\eqref{eq:S21_coupled} is required to derive the resonator response and reproduce the observations. Following the derivation from Ref. \cite{s2016_bur}, we first describe the interaction Hamiltonian between the multi-level QDs and the microwave photons, where the resonator effectively couples to the DQD detuning $\varepsilon$ as shown below Supplementary Eq.~\eqref{eq:SE4.1}. It should be noted that the same interaction Hamiltonian can be utilized for both the even and the odd configuration reported in Fig. 6, because $\varepsilon$ couples to the Hamiltonian via $\tau_z$, which is the dipole moment operator in the position basis $[|\mathrm{L}_\text{g}\rangle, |\mathrm{L}_\text{e}\rangle, |\mathrm{R}_\text{g}\rangle, |\mathrm{R}_\text{e}\rangle]$, where L (R) denotes the charge state with the excess hole in the left (right) QD (see Methods).

\begin{equation} \label{eq:SE4.1}
\begin{split}
H_\text{int} & = g_0 \tau_\text{z} (a + a^{\dagger}) \\
& =g_0 
\begin{bmatrix}
1 & 0 & 0 & 0 \\
0 & \eta_\text{L} & 0 & 0 \\
0 & 0 & -1 & 0 \\
0 & 0 & 0 & -\eta_\text{R} \\
\end{bmatrix}
(a + a^{\dagger}). 
\end{split}
\end{equation} 
Here, $\eta_\text{L}$ and $\eta_\text{R}$ account for the different lever-arms of the excited orbital states as discussed in the Methods section.
Based on $H_\text{int}$ we use the input-output theory to simulate the normalized feedline transmission amplitude $|A/A_0|^2$ shown in Fig. 5b, and in Supplementary Figure~\ref{fig:Extended data even odd spectrum} for both odd and even cases. The transmission of the hanged-style resonator coupled to multi-level QDs can be written as:

\begin{equation} \label{eq:SE4.2}
S_\text{21} = |A/A_0| =\frac{\Delta_\mathrm{r} - i\kappa_\mathrm{int}/2 + g_0 \sum_{n, m} d_\mathrm{nm} \chi_\mathrm{nm}}
{\Delta_\mathrm{r} - i\kappa/2 + g_0 \sum_{n, m} d_\mathrm{nm} \chi_\mathrm{nm}},
\end{equation} 
where the dipole moment operator $d$ and the charge susceptibility $\chi$ are matrices to account for different state transitions. The dipole moment operator $d$ can be evaluated by transforming $\tau_z$ from the position basis $[|\mathrm{L}_\text{g}\rangle, |\mathrm{L}_\text{e}\rangle, |\mathrm{R}_\text{g}\rangle, |\mathrm{R}_\text{e}\rangle]$ to the `qudit' basis which diagonalizes the multi-level Hamiltonian (shown in Methods):
\begin{equation}
    H_\mathrm{SCS, diag}=U_\mathrm{0}H_\mathrm{SCS}U_\mathrm{0}^{\dagger}=\sum_{m,n=0}^3 E_\mathrm{n}\vert n\rangle\langle n\vert,
\end{equation}
\begin{equation} \label{eq:dipole_moment}
    d = U_\mathrm{0}\tau_\mathrm{z}U_\mathrm{0}^{\dagger} = \sum_{m,n=0}^3 d_{\mathrm{mn}}\vert m\rangle\langle n\vert,
\end{equation}
where $U_\mathrm{0}$ is a unitary transformation which diagonalizes $H_\mathrm{SCS}$, $E_\mathrm{n}$ ($|n\rangle$) are the eigenenergies (eigenstates) of $H_\mathrm{SCS, diag}$, $d_\mathrm{nm} = d_\mathrm{mn}^*$  are the matrix elements of $d$.

The matrix elements of $\chi$ can be written as:
\begin{equation} \label{eq:SE4.3}
\chi_\mathrm{nm} = \frac{g_0 d_\mathrm{nm} (\rho_\mathrm{m} - \rho_\mathrm{n})}
{-(E_\mathrm{n}/\hbar - E_\mathrm{m}/\hbar - \omega_\mathrm{d}) + i\gamma_\mathrm{nm}},
\end{equation} 
with $\gamma_\mathrm{nm} = \Gamma_\mathrm{nm}^\varphi + \Gamma_\mathrm{nm}^\mathrm{r}/2$ representing the total loss rate given by the dephasing (relaxation) rate $\Gamma_\mathrm{nm}^\varphi$ ($\Gamma_\mathrm{nm}^\mathrm{r}$) between n and m states. In Supplementary Eq.~\eqref{eq:SE4.3}, and $\rho_i \propto e^{-E_i/(k_\mathrm{B}T)}$ is the normalized Boltzmann distribution. Since the quantum states are far detuned below the Fermi-level, we assume $T = 10$ mK $\ll$ $T_\mathrm{e}$.

To treat the decoherence matrix $\gamma$, we introduce the noise Hamiltonian $H_n$ Supplementary Eq.~\eqref{eq:SE4.4} in the position basis \cite{s2017_stoa}:
\begin{equation} \label{eq:SE4.4}
\begin{aligned}
H_n & = N_\varepsilon + \sum_{i,j} N_{ij} \\
N_\varepsilon & = \xi_{\varepsilon} \tau_\mathrm{z} \\
N_{ij} & = \xi_{ij}(\vert i\rangle\langle j\vert + \vert j\rangle\langle i\vert).
\end{aligned}
\end{equation}
Here $N_\varepsilon$ ($N_{ij}$) illustrates the noise Hamiltonian with $\xi_{\varepsilon}$ ($\xi_{ij}$) representing randomly fluctuating noise with the corresponding power spectra  $P_\varepsilon(\omega)$ ($P_{ij}(\omega)$). 
For instance, when $\varepsilon \gg 0$, the qudit basis describes decoupled QDs in the position basis, and $P_{ij}(\omega_\mathrm{i} - \omega_\mathrm{j}) \propto \Gamma_{ij}^{\mathrm{b,r}}$ holds \cite{s2017_stoa}, where $\omega_\mathrm{i} = E_\mathrm{i}/\hbar$  and $\Gamma_{ij}^{\mathrm{b,r}}$ is the relaxation rate of the bare quantum state from $|i\rangle$ to $|j\rangle$.
In the qudit basis, we interpret $\bar{N}_{nm}^2 \propto \Gamma_{nm}^\mathrm{r}$ and $(\bar{N}_{nn} - \bar{N}_{mm})^2 \propto \Gamma_\mathrm{nm}^\varphi$, with $\bar{N}$ representing the noise Hamiltonian in the qudit basis.
We also assume that the low-frequency noise affects the detuning noise, i.e. $P_\varepsilon(0) \sim \Gamma_\varepsilon$ \cite{s2017_stoa}.
For the simulations shown in Fig. 5b and Supplementary Figure~\ref{fig:Extended data even odd spectrum}, we empirically use the decoherence parameters of the individual QDs in the position basis shown in Supplementary Table~\ref{table:decoherence_matrix}. We note, however, that experiments such as the free-induction-decay \cite{s2013_dia} or relaxation time measurement \cite{s2020_law} are required to accurately determine the decoherence parameters.

\begin{table}[h!]
\centering
\begin{tabularx}{0.8\linewidth}{ >{\centering\arraybackslash}p{2cm} >{\centering\arraybackslash}p{2cm} >{\centering\arraybackslash}X  >{\centering\arraybackslash}X
}
\toprule
Parameter & Fig. 5 & Supplementary Figure~\ref{fig:Extended data even odd spectrum}a& Supplementary Figure~\ref{fig:Extended data even odd spectrum}b\\ 
& (MHz) & (MHz) & (MHz) \\
 \midrule
$\Gamma_\varepsilon/2\pi$  & 180 & 160 & 180 \\ 
$\Gamma_{12}^{\mathrm{b,r}}/2\pi$ & 1 & 1 & 1 \\
$\Gamma_{34}^{\mathrm{b,r}}/2\pi$ & 1 & 1 & 1 \\
$\Gamma_{13}^{\mathrm{b,r}}/2\pi$ & 1 & 1 & 1 \\
$\Gamma_{14}^{\mathrm{b,r}}/2\pi$ & 1 & 1 & 1 \\
$\Gamma_{24}^{\mathrm{b,r}}/2\pi$ & 30 & 10 & 1 \\
$\Gamma_{23}^{\mathrm{b,r}}/2\pi$ & 1 & 90 & 1 \\
\bottomrule
\end{tabularx}
\caption{Qudit decoherence matrix utilized for the numerical simulations shown in Fig. 5 and Supplementary Figure~\ref{fig:Extended data even odd spectrum}.}
\label{table:decoherence_matrix}
\end{table}

\newpage
\section{Spectroscopies of the strongly-correlated states}\label{sec:WM_details}

\begin{figure}[H]
     \centering
     \includegraphics[width=90mm]{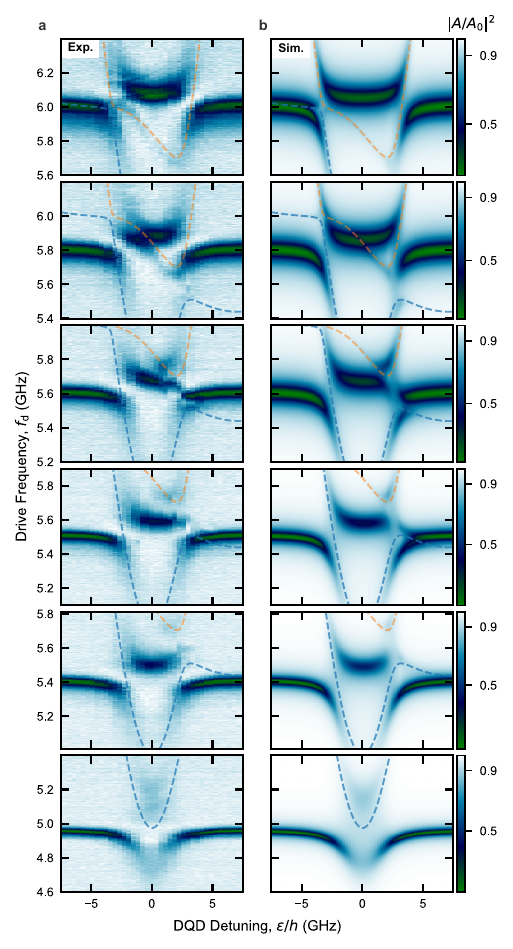}
     \caption{
     \textbf{Detailed spectroscopy of the SCS.}
     \textbf{a,} Detailed normalized amplitude of feedline transmission $|A/A_\mathrm{0}|^2$ as a function of drive frequency $f_\mathrm{d}$ and DQD detuning $\varepsilon$, obtained in correspondence of six different resonator frequencies $f_\mathrm{r}$ and for the same inter-dot transition shown in Fig.~5.
     Blue and orange dashed lines (identical to the ones shown in Fig.~5) correspond to the first and second excited spectrum in the presence of the quenched SCS. 
     \textbf{b,} Simulation of $|A/A_\mathrm{0}|^2$ obtained by the multi-level DQD input-output theory. The parameters used for the simulations are reported in Supplementary Table~\ref{table:fig5_parameters}. 
     }
     \label{fig:Extended data WM}
\end{figure}

\begin{table}[h!]
\centering
\begin{tabularx}{0.4\linewidth}{ >{\centering\arraybackslash}X >{\centering\arraybackslash}X 
}
\toprule
Parameter & Fig.~5 ($h\cdot$GHz) \\ [0.5ex]
 \midrule
$\Delta_\mathrm{L}$  & 5.40  \\ 
$\Delta_\mathrm{R}$ & 4.73 \\
$t_{11}$ & 2.49 \\
$t_{12}$ & 0.21 \\
$t_{21}$ & 0.11  \\
$t_{22}$ & 1.69 \\
$\hbar g_{0}$ & 0.22 \\
\bottomrule
\end{tabularx}
\caption{SCS Hamiltonian parameters (see Eq.~(2) in Methods) reproducing the experimental data reported in Fig.~5a, and Supplementary Figure~\ref{fig:Extended data WM}. The resonator parameters, including $\kappa_\mathrm{int}$ and $\kappa$, are extracted from Fig.~1g in correspondence of the $f_\mathrm{r}$ used in the different panels.}
\label{table:fig5_parameters}
\end{table}

\begin{figure}[H]
     \centering
     \includegraphics[width=90mm]{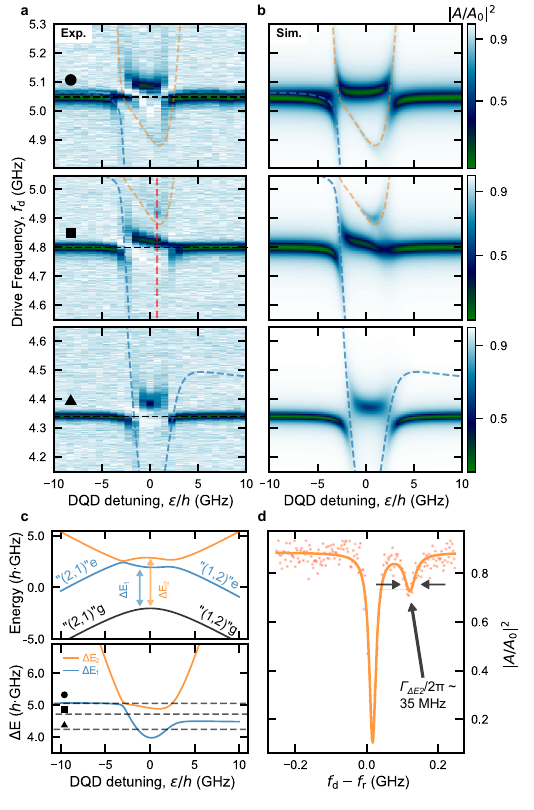}
     \caption{
     \textbf{SCS spectroscopy in a different electrostatic configuration.}
     \textbf{a,} Normalized amplitude of feedline transmission $|A/A_\mathrm{0}|^2$ as a function of drive frequency $f_\mathrm{d}$ and DQD detuning $\varepsilon$ for three different resonator frequencies $f_\mathrm{r}$. An inter-dot transition different from the one investigated in Fig.~5 is studied here. 
     \textbf{b,} Simulated $|A/A_\mathrm{0}|^2$ for the three different $f_\mathrm{r}$ in \textbf{a}, using a generalized input-output theory for multi-level DQD systems (see Methods, and Supplementary Note~\ref{sec:io_wigner}). The relevant parameters are shown in Supplementary Table~\ref{table:another_WM_Parameters}.
     \textbf{c,} Energy-level diagram (top panel) and excitation energy $\Delta_E$ (bottom panel) calculated with the 4 $\times$ 4 Hamiltonian in Methods, and used for the input-output simulation in \textbf{b}. Bottom panel: Blue (orange) curve corresponds to the energy spitting $\Delta E_\text{1}$ ($\Delta E_\text{2}$) between the first (second) excited state and the ground state shown in the upper panel. $\Delta E_\text{1}$ ($\Delta E_\text{2}$) spectrum is superimposed to \textbf{a} and \textbf{b} in blue (orange) dashed line.
     \textbf{d,} Line-cut along the red dashed line shown in the middle panel of \textbf{a,}. A fit to a Lorentizan model results in $\Gamma_{\Delta E_2}/2\pi \sim 35$~MHz. 
     }
     \label{fig:Extended data another WM}
\end{figure}

\begin{table}[h!]
\centering
\begin{tabularx}{0.4\linewidth}{ >{\centering\arraybackslash}X >{\centering\arraybackslash}X }
\toprule
Parameter & Value ($h\cdot$GHz) \\ [0.5ex]
\midrule
$\Delta_\mathrm{L}$  & 5.03  \\ 
$\Delta_\mathrm{R}$ & 4.45 \\
$t_{11}$ & 2.00 \\
$t_{12}$ & 0.50 \\
$t_{21}$ & 0.22 \\
$t_{22}$ & 1.90 \\
$\hbar g_{0}$ & 0.13 \\
\bottomrule
\end{tabularx}
\caption{SCS Hamiltonian parameters (see Eq.~(2) in Methods) reproducing the experimental data reported in Supplementary Figure~\ref{fig:Extended data another WM}. The resonator parameters, including $\kappa_\mathrm{int}$ and $\kappa$, are extracted from Fig.~1g in correspondence to the $f_\mathrm{r}$ used in the different panels.}
\label{table:another_WM_Parameters}
\end{table}

\newpage
\section{DC transport measurements in the even and odd hole configurations}\label{sec:even_odd}

\begin{figure}[H]
     \centering
     \includegraphics[width=1\textwidth]{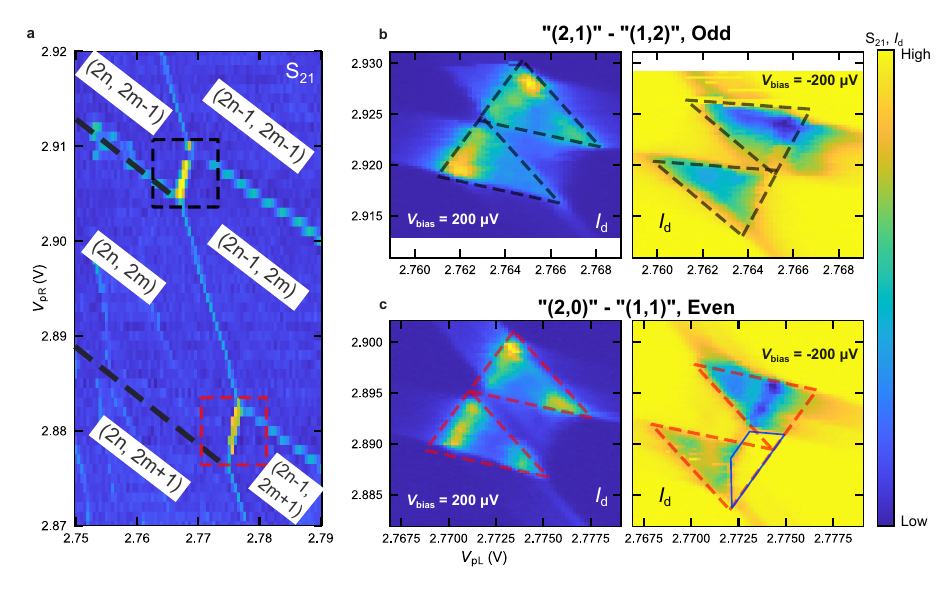}
     \caption{
     \textbf{DC bias triangle measurements in the even and odd total hole number configurations. ~~~} 
     \textbf{a,} Stability diagram spanned by $V_\mathrm{pL}$ and $V_\mathrm{pR}$ showing two adjacent inter-dot transitions. Amplitude of the feedline transmisson $S_{21}$ is recorded. (p, q) denotes the relevant charge configuration where p (q) is the hole number in the left (right) QD.
     \textbf{b,} Bias-triangle measurements at the (2n, 2m-1) $\leftrightarrow$ (2n-1, 2m) ($``(2, 1)" \leftrightarrow ``(1, 2)"$) inter-dot transition (black-dashed box in \textbf{a}), corresponding to the one investigated in Fig. 6a. The dc-current through the DQD, $I_\mathrm{d}$, is recorded. Left (right) panel shows the bias-triangle measured with $V_\mathrm{bias}$ = 200 $\mu$V (-200 $\mu$V). The black-dashed triangles with the same shape and size are superimposed to the both panels, showing no indication of the Pauli spin blockade (PSB). 
     \textbf{c,} The same measurement as in \textbf{b} at the (2n, 2m) $\leftrightarrow$ (2n-1, 2m+1) ($``(2, 0)" \leftrightarrow ``(1, 1)"$) inter-dot transition (red-dashed box in \textbf{a}) corresponding to the one investigated in Fig. 6b. The red-dashed triangles with the same shape and size are superimposed to the both panels, where in the right panel, the possible indication of the PSB is denoted by the blue trapezoid.
     }
     \label{fig:Fig_EvenOdd_Stab}
\end{figure}


Two adjacent inter-dot transitions corresponding to those investigated with the resonator in Fig. 6 were characterized through dc-transport measurements. Supplementary Figure~\ref{fig:Fig_EvenOdd_Stab}a displays these transitions in terms of the feedline transmission amplitude, $S_{21}$, as a function of $V_\mathrm{pR}$ and $V_\mathrm{pL}$. 
When a finite bias voltage, $V_\mathrm{bias}$, is applied at the source ohmic contact, bias triangles near the triple points in the charge stability diagrams emerge \cite{s2005_joh, s2007_han}, as shown in Supplementary Figure~\ref{fig:Fig_EvenOdd_Stab}b. The left panel represents the bias triangles measured under $V_\mathrm{bias} = + 200 ~\mu$V, while the right panel corresponds to $V_\mathrm{bias} = - 200 ~\mu$V, both at the (2n, 2m-1) $\leftrightarrow$ (2n-1, 2m) ($``(2, 1)" \leftrightarrow ``(1, 2)"$) inter-dot transition, as denoted by the odd, black dashed box in Supplementary Figure~\ref{fig:Fig_EvenOdd_Stab}a. To facilitate a comparison of the bias triangles, superimposed black dashed triangles of the same size but with different orientations were included in Supplementary Figure~\ref{fig:Fig_EvenOdd_Stab}b. When the size of the bias triangles closely matches, it implies no indication of the Pauli spin blockade (PSB).

In contrast, Supplementary Figure~\ref{fig:Fig_EvenOdd_Stab}c depicts the same measurement as in Supplementary Figure~\ref{fig:Fig_EvenOdd_Stab}b but at the (2n, 2m) $\leftrightarrow$ (2n-1, 2m+1) ($``(2, 0)" \leftrightarrow ``(1, 1)"$) inter-dot transition (even configuration, red dashed box in Supplementary Figure~\ref{fig:Fig_EvenOdd_Stab}a), corresponding to the one investigated in Fig. 6b. 
In this case, representing the even scenario, the size of the bias triangles (red dashed triangles in Supplementary Figure~\ref{fig:Fig_EvenOdd_Stab}c) measured with the opposite polarity of $V_\mathrm{bias}$ does not match, revealing a region where the current is blocked (blue trapezoid in the right panel). While more detailed studies, including the introduction of a finite magnetic field \cite{s2008_sha}, are required to undeniably confirm this observation, the observed current blockade presents a potential indication of the Pauli spin blockade (PSB). The PSB is expected for inter-dot transitions with a total even number of particles \cite{s2005_joh}. From the width of the PSB region in the DQD bias triangles, we extract an orbital splitting of $\sim 7.8 \pm 1.2 $ $h\cdot$GHz (\add{where the error is determined by the size of voltage step utilized in Supplementary Figure~\ref{fig:Fig_EvenOdd_Stab}c}), which aligns with the estimated $\Delta_\mathrm{L} \sim 5.46 ~ h\cdot$GHz in Fig. 6a, providing further support for the presence of strongly-correlated states in this study.

\newpage

\section{Simulations of spectroscopies in Fig. 6}\label{sec:even_odd_spectroscopy}

\begin{figure}[H]
     \centering
     \includegraphics[width=90mm]{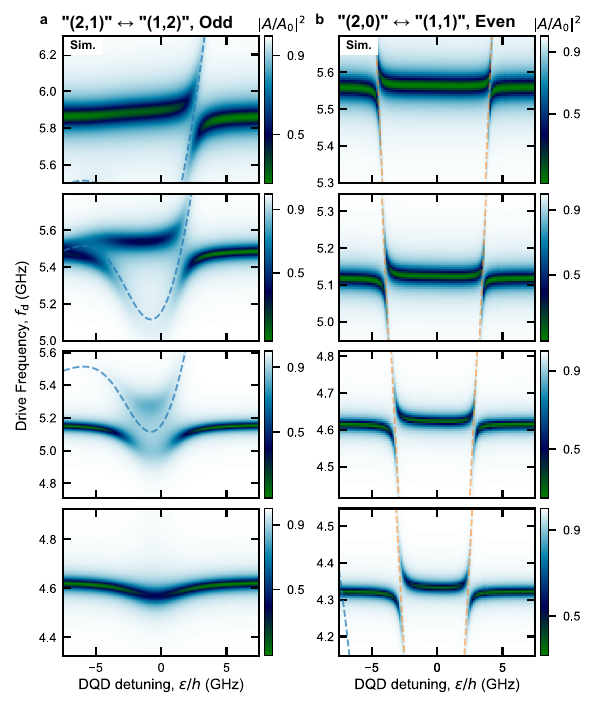}
     \caption{
     \textbf{Simulation of the resonator spectra shown in Fig.~6.}
     Simulated normalized amplitude of feedline transmission $|A/A_\mathrm{0}|^2$ of the corresponding measurements in Fig.~6, using an extended master equation model of the multi-level DQD system(see Methods and Supplementary Note~\ref{sec:io_wigner}) for an odd (\textbf{a}, see Fig.~6a) and an even (\textbf{b}, see Fig.~6b) DQD configuration, respectively. Relevant parameters for the simulations are reported in Supplementary Table~\ref{table:even_odd_parameters}, which are also used for calculating the energy level diagram and the excitation energy spectra shown in Fig.~6e and f.
     The observed asymmetry in $\varepsilon$ of Fig.~6a, as already in Fig.~5, signals the presence of an additional excited state associated with a SCS. This results in an excitation energy around $\sim 5 ~h\cdot$GHz for the left QD, while the excited state of the right QD cannot be resolved up to $f_{r}\sim$ 6.3~GHz. 
     The observed asymmetry in $\varepsilon$, as shown in Fig.~6a, signals the presence of an additional excited state associated with a SCS around $\sim 5 ~h\cdot$GHz in the left QD. 
     }
     \label{fig:Extended data even odd spectrum}
\end{figure}

\begin{table}[h!]
\centering
\begin{tabularx}{0.6\linewidth}{ >{\centering\arraybackslash}X >{\centering\arraybackslash}X >
{\centering\arraybackslash}X 
}
\toprule
Parameter &  Fig.~6a, Odd ($h\cdot$GHz) & Fig.~6b, Even ($h\cdot$GHz)\\ [0.5ex]
\midrule
$\Delta_\mathrm{L}$ & 5.48 & 5.48 \\ 
$\Delta_\mathrm{R}$ & 6.50$^*$ & 0.00 \\
$t_{11}$ & 2.65 & 1.60 \\
$t_{12}$ & 0.30$^*$ & 0.00 \\
$t_{21}$ & 1.25 & 0.00 \\
$t_{22}$ & 1.70$^*$ & 2.20 \\
$\hbar g_{0}$ & 0.19 & 0.12 \\
 \bottomrule
\end{tabularx}
\caption{SCS Hamiltonian parameters reproducing the experimental data reported in Fig.~6a, and b. The resonator parameters including $\kappa_\mathrm{int}$ and $\kappa$ are extracted from Fig.~1g in correspondence to the $f_\mathrm{r}$ used in the different panels. For this specific DQD configuration, the parameters indicated by $^*$ are the arbitrary values which cannot be determined from the measurement in Fig.~6a and do not impact the results of the simulations (see caption of Supplementary Figure~\ref{fig:Extended data even odd spectrum}).} 
\label{table:even_odd_parameters}
\end{table}

\newpage


\end{document}